\documentclass[usenatbib,fleqn]{mnras}
\usepackage{graphicx,hyperref,amsmath,amssymb,xspace}
\usepackage[dvipsnames]{xcolor}

\title[Sgr, LMC, and the Milky Way]{Tango for three: Sagittarius, LMC, and the Milky Way}

\author[Vasiliev, Belokurov \& Erkal]{
Eugene Vasiliev$^{1,2}$\thanks{E-mail: eugvas@lpi.ru}, Vasily Belokurov$^1$, Denis Erkal$^3$\\
$^1$Institute of Astronomy, Madingley road, Cambridge, CB3 0HA, UK\\
$^2$Lebedev Physical Institute, Leninsky prospekt 53, Moscow, 119991, Russia\\
$^3$Department of Physics, University of Surrey, Guildford GU2 7XH, UK}

\newcommand{\Gaia}{\textit{Gaia}\xspace}
\newcommand{\kms}{km\:s$^{-1}$\xspace}
\newcommand{\masyr}{mas\:yr$^{-1}$\xspace}
\newcommand{\Beta}{\mathrm{B}}
\defcitealias{Law2010a}{LM10}
\defcitealias{Vasiliev2020}{VB20}

\begin{document}
\date{Accepted 2020 November 20. Received 2020 November 16; in original form 2020 September 22}
\pagerange{2279--2304}\volume{501}\pubyear{2021}
\setcounter{page}{2279}
\maketitle

\begin{abstract}
We assemble a catalogue of candidate Sagittarius stream members with 5d and 6d phase-space information, using astrometric data from \Gaia DR2, distances estimated from RR Lyrae stars, and line-of-sight velocities from various spectroscopic surveys. We find a clear misalignment between the stream track and the direction of the reflex-corrected proper motions in the leading arm of the stream, which we interpret as a signature of a time-dependent perturbation of the gravitational potential. A likely cause of this perturbation is the recent passage of the most massive Milky Way satellite -- the Large Magellanic Cloud (LMC). We develop novel methods for simulating the Sagittarius stream in the presence of the LMC, using specially tailored $N$-body simulations and a flexible parametrization of the Milky Way halo density profile. We find that while models without the LMC can fit most stream features rather well, they fail to reproduce the misalignment and overestimate the distance to the leading arm apocentre. On the other hand, models with an LMC mass in the range $(1.3\pm0.3)\times10^{11}\,M_\odot$ rectify these deficiencies. We demonstrate that the stream can not be modelled adequately in a static Milky Way. Instead, our Galaxy is required to lurch toward the massive in-falling Cloud, giving the Sgr stream its peculiar shape and kinematics. By exploring the parameter space of Milky Way potentials, we determine the enclosed mass within 100~kpc to be $(5.6\pm0.4)\times10^{11}\,M_\odot$, and the virial mass to be $(9.0\pm1.3)\times10^{11}\,M_\odot$, and find tentative evidence for a radially-varying shape and orientation of the Galactic halo.
\end{abstract}


\section{Introduction}   \label{sec:intro}

The Sagittarius dwarf galaxy \citep[Sgr,][]{Ibata1994}, one of the closest Milky Way satellites, is undergoing tidal disruption, producing a spectacular stream of debris spanning the entire sky. The first glimpse of the stream's enormous dimensions was revealed with the advent of the 2MASS survey: \citet{Majewski2003} discovered the two arms of the stream -- the leading in the northern Galactic hemisphere and the trailing in the southern hemisphere. The large spatial extent of the stream makes it a powerful observational probe of the Milky Way gravitational field. Numerous studies attempted to constrain the matter distribution in the Milky Way by fitting the positions, line-of-sight velocities and distances to the stream stars. Usually the parametric description of the dark halo's potential is optimized to fit the models to the observations.  However, it soon became clear that the Sgr stream presents more puzzles than expected, as the different subsets of data drove the models into incompatible regions of the parameter space. For instance, the velocities of the leading arm favour a prolate halo \citep{Helmi2004, Law2005} while the spatial appearance is better described by an oblate halo \citep{Johnston2005}.

A possible solution to this conundrum was suggested in \citet{Law2010a}, hereafter LM10 (and confirmed by \citealt{Deg2013}), who took advantage of the large number of line-of-sight velocities along both arms \citep[see][]{Majewski2004, Law2005}, as well as accurate measurements of the leading tail structure provided by \citet{Belokurov2006}, to explore a more general class of triaxial potentials with an arbitrary orientation of the principal axes of the halo, and presented an $N$-body model that successfully reproduced most of the then-known stream properties. This model requires a slightly triaxial (nearly oblate) potential, whose minor axis, however, lies in the Galactic plane and points roughly towards the Sgr remnant, and the intermediate axis is perpendicular to the Milky Way disc. However, as argued by \citet{Debattista2013}, such a configuration is unstable dynamically and very unlikely to arise in nature.

Intriguingly, the minor axis of the Milky Way halo in the \citetalias{Law2010a} model is also approximately aligned with the orbital pole of its largest satellite -- the Large Magellanic Cloud (LMC). This coincidence, already acknowledged in that paper, may indicate that the peculiar shape and orientation of the halo simply offer an effective description for the gravitational influence of the LMC. Various pieces of evidence suggest that the LMC has mass $\gtrsim 10^{11}\,M_\odot$, i.e. a significant fraction of the Milky Way mass, and is on its first approach to the Milky Way \citep[e.g.,][]{Besla2007}. Recognizing this fact, \citet{VeraCiro2013} constructed relatively simple models for the stream in the presence of a moving LMC, and found that it does have a substantial effect on the inferred properties of the Milky Way potential. Their study, however, missed another important piece of physics -- the reflex motion of the Milky Way centre caused by the gravitational tug of the LMC. \citet{Gomez2015} took this factor into account in a suite of $N$-body simulations and confirmed the significant role that the LMC can play in shaping the Sgr's tidal tails, but did not attempt to fit the stream's properties in detail.

Recently, multiple lines of evidence have converged to support the heavy LMC hypothesis, as suggested by stellar mass--halo mass relation \citep[see e.g.][]{Guo2010} and explored for the first time in \citet{Kallivayalil2013}. For example, the so-called timing argument was used by \citet{Penarrubia2016} to arrive at the mass of $M_{\rm LMC} \sim (2.5\pm1)\times10^{11} M_{\odot}$. The recently discovered faint satellites of the LMC \citep[see e.g.][]{Jethwa2016,Sales2017,Kallivayalil2018} can also be used to place a limit on the Cloud's mass, as has been done by \citet{Erkal2020}, who inferred $M_{\rm LMC}\gtrsim 1.2\times10^{11} M_{\odot}$. The influence of a massive LMC on the kinematics of both the smooth Milky Way halo and stellar streams has received renewed attention. The enormous Cloud can re-arrange the Milky Way, decoupling the outer halo of the Galaxy from its inner regions and spawning a trailing wake \citep[see e.g.][]{GaravitoCamargo2019,GaravitoCamargo2020,Petersen2020}. As a consequence, the kinematics of the halo tracers (both stars and satellites) acquire specific asymmetry patterns, which have been detected recently by \citet{Erkal2020b}; these distortions need to be compensated when applying traditional dynamical modelling methods such as Jeans equations \citep[e.g.][]{Deason2020}, otherwise the Galactic mass estimates relying on the assumption of dynamical equilibrium would be biased \citep{Erkal2020a}. The most striking proof of the LMC's action is presented in \citet{Erkal2019}, who demonstrated that the long and thin Orphan stream can be successfully reproduced only by models that include a perturbation from the LMC, estimating the mass of the latter to be $\sim (1.4\pm0.3)\times10^{11}\,M_\odot$. The main piece of evidence is the misalignment of the reflex-corrected proper motion of stream stars with the stream itself (found by \citealt{Koposov2019}), which is not possible in a static potential. As will be shown in this Paper, a similar misalignment is also apparent in the leading arm of the Sgr stream.

Over the years, the observational coverage of the Sgr stream gradually increased and presented even more puzzles, such as the discovery of fainter parallel streams in the leading arm \citep{Belokurov2006} and in the trailing arm \citep{Koposov2012}, and the factor of two difference between the apocentre distances of the leading (50~kpc) and the trailing (100~kpc) arm based on Blue Horizontal Branch stars and RR Lyrae (e.g., \citealt{Belokurov2014}, \citealt{Hernitschek2017}). The full extent of the trailing arm was not apparent at the time of \citetalias{Law2010a} (although see \citealt{Newberg2003} for an early evidence), and their model does not match its apocentre distance by a large margin. More recent modelling efforts (e.g., \citealt{Gibbons2014}, \citealt{Dierickx2017}, \citealt{Fardal2019}) improve the fit to the structural properties of the stream, but still fall short of reproducing all of its observed features.

The second data release (DR2) of the \Gaia mission \citep{Brown2018} opened a new all-sky window on the Milky Way and its surroundings. Although limited information about proper motions in the Sgr stream was available before \citep[see e.g.][]{Carlin2012,Koposov2013,Sohn2015}, the \Gaia data expanded the available sample of stars with proper motions by orders of magnitude. The first wide-area proper motion study of the stream is presented in \citet{Deason2017}, who used a combination of the \Gaia DR1 data and the re-calibrated SDSS astrometry. Subsequently, \citet{Antoja2020} used \Gaia DR2 to assemble an all-sky selection of possible Sgr members based just on the two proper motion components, while \citet{Ramos2020} augmented it with the distance information based on RR Lyrae stars from the \Gaia catalogue. Independently, \citet{Ibata2020} constructed another sample of $\sim2.6\times10^5$ candidate members of the Sgr stream, using the \textsc{STREAMFINDER} algorithm \citep{Malhan2018}. This sample includes $3\,500$ RR Lyrae and $\sim3\,000$ giants with line-of-sight velocity information from various spectroscopic surveys, thus providing a 6d view on the stream. These recent studies have compared their observational samples with the \citetalias{Law2010a} model and found reasonable agreement in most parts of the stream, but with important deviations, especially in the distant portions of the trailing arm. Finally, in \citet{Vasiliev2020}, hereafter VB20, we constructed a detailed map of the Sgr remnant, based on \Gaia DR2 proper motions and line-of-sight velocities from extant spectroscopic datasets, and performed a large suite of $N$-body simulations of a disrupting Sgr galaxy, aiming at reproducing the properties of the remnant. While we were able to delineate a range of models that satisfy these observational constraints, these simulations were not designed to fit the stream (and indeed they did not match its leading arm geometry). In the present paper we address this deficiency.

The aim of this paper is twofold.  First, we construct a sample of red giant stars in both arms of the stream with five- and six-dimensional phase-space information, combining the proper motion measurements from \Gaia, distances from RR Lyrae, and line-of-sight velocities from various spectroscopic surveys. This catalogue, which we make publicly available, contains $\sim 55\,000$ stars with an estimated contamination fraction of only a few percent, including $\sim 4\,500$ stars with six-dimensional phase-space information. We then demonstrate that the leading arm of the stream shows a clear misalignment between the stream track and reflex-corrected proper motions, which is a smoking-gun signature of a time-dependent perturbation to the stream \citep[e.g.][]{Erkal2019,Shipp2019}. In the present study, we assume that this perturbation is caused by the LMC flyby. Second, we develop new techniques for constructing $N$-body models of the disrupting Sgr galaxy and its stream in the presence of the LMC. We conduct an extensive search through the parameter space of Milky Way potentials and identify the values that produce a realistically looking Sgr stream. Although models without the LMC can match most observed properties reasonably well, they fail in the crucial test -- the misalignment between the reflex-corrected proper motions and the stream track. However, models with the LMC mass around $\sim1.5\times10^{11}\,M_\odot$ are able to qualitatively reproduce this misalignment, as well as adequately fit most other properties of the stream.

The paper is organized as follows. In Section~\ref{sec:data} we construct the selection of high-confidence stream candidates. In Section~\ref{sec:method} we review various modelling techniques and describe our approach. Then in Section~\ref{sec:fit_results} we present the results of model fits with and without the LMC and compare them with various features of the Sgr stream. We discuss the physical mechanisms of the three-galaxy interaction and the constraints on the Milky Way potential provided by these models in Section~\ref{sec:discussion}, and summarize our findings in Section~\ref{sec:conclusions}.

\section{Observational data}   \label{sec:data}

\begin{figure*}
\includegraphics[width=\textwidth]{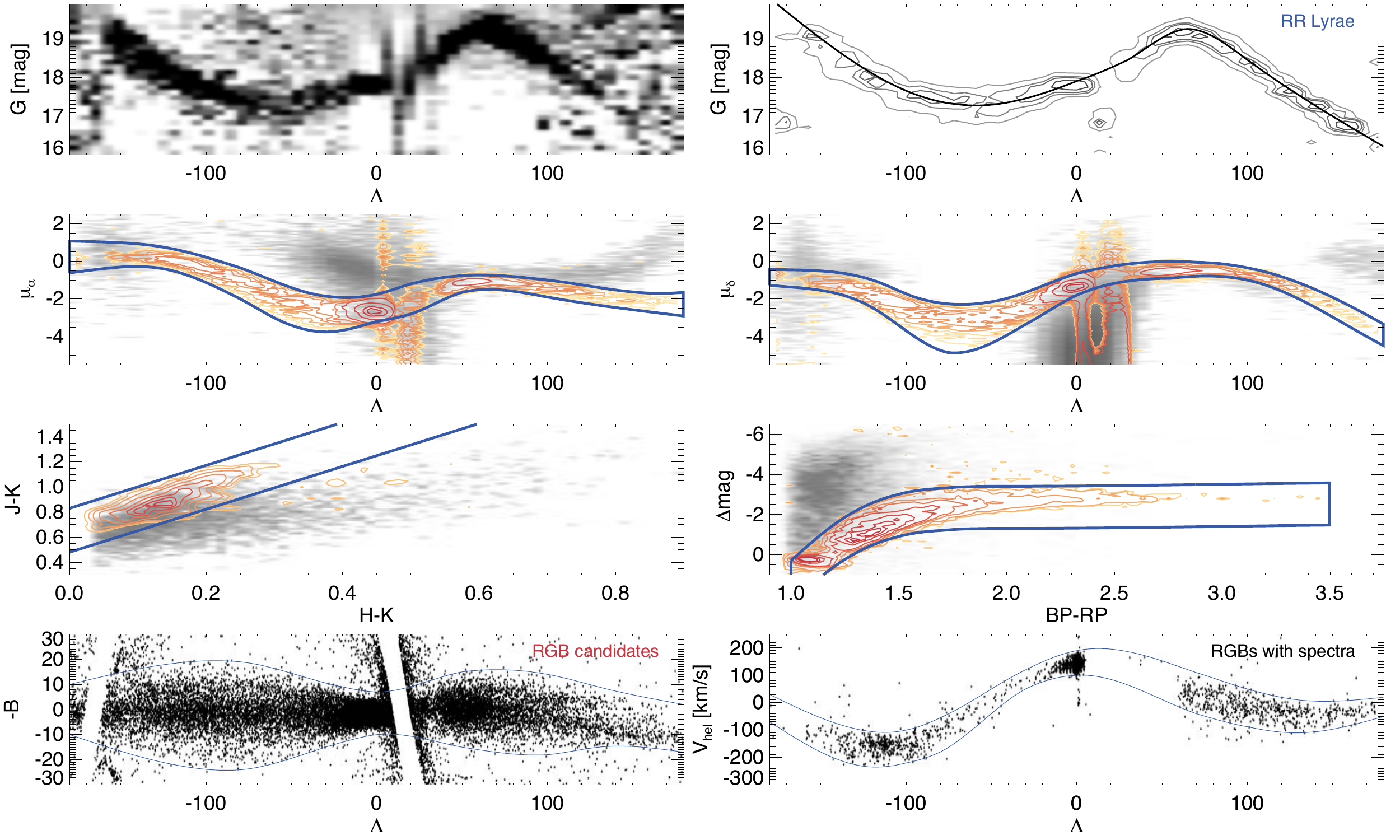}
\caption{Sagittarius stream selection with \Gaia DR2 and 2MASS. 
{\it 1st row from the top, left panel:} Density difference between \Gaia RR Lyrae inside the stream ($|\Beta|<10^{\circ}$) and outside the stream ($20^{\circ}<|\Beta|<40^{\circ}$)) in the plane of apparent magnitude $G_{\rm RRL}$ and stream longitude $\Lambda$. Dark, snaking ribbon corresponds to the overdensity of the stream RR Lyrae. 
{\it 1st row, right:} Density contours of the stream RR Lyrae in the same plane as in the previous panel. The stream RR Lyrae members are selected using the $\Beta$ cut as well as the proper motion cuts outlined below. Thick solid curve is the distance ridgeline used to describe the variation in the stream proximity to the observer as a function of $\Lambda$. 
{\it 2nd and 3rd row:} Density of the RGB stream member candidates (yellow--red contours) overplotted on top of the foreground density (greyscale) in proper motion components $\mu_{\alpha}, \mu_{\delta}$, 2MASS colour $J-K, H-K$, and \Gaia colour and absolute magnitude $BP-RP, M_G$. Note that these stars are preselected using criteria involving parallax, apparent magnitude $G$ and colours $H-K$ and $BP-RP$. In each of the four panels, the stream and the foreground are filtered with a combination of cuts shown as thick blue lines in the surrounding panels but excludes the selection shown in the panel itself (see Section~\ref{sec:rgb_sel} for detailed explanation). 
{\it Bottom row, left:} Candidate RGB stream members selected using a combination of the colour, magnitude and proper motion cuts illustrated above. 
{\it Bottom row, right:} Extant heliocentric line-of-sight velocity measurements for a sample of candidate RGB members selected as shown in the previous panel.
} \label{fig:observational_selection}
\end{figure*}

\subsection{Coordinates}

We define the \textit{right-handed} celestial coordinate system $\Lambda,\Beta$ that differs from the one used in \citet{Majewski2003} by the sign of $\Lambda$: it increases towards the leading arm, and is zero at the centre of the Sgr remnant. Rather annoyingly, the other coordinate ($\Beta$) is \textit{not} centered on the remnant: the latter has $\Beta_0\approx 1.5^\circ$; we follow this convention to match previous studies. \citet{Belokurov2014} introduced an alternative definition of the coordinate system, which flips the signs of both $\Lambda$ and $\Beta$. Both of these choices create a left-handed coordinate system, unlike the other commonly used celestial coordinates (ICRS or Galactic), which we believe to be confusing. Our convention uses $\Lambda$ from \citet{Belokurov2014} and $\Beta$ from \citet{Majewski2003}, thus the leading arm has positive $\Lambda$, and the orbital pole of the Sgr plane has $\Beta=90^\circ$ ($l=273.75^\circ$, $b=-13.46^\circ$ in Galactic coordinates, or $\alpha=123.65^\circ$, $\delta=-53.52^\circ$ in ICRS).

We use the standard right-handed Galactocentric coordinate system as defined in \textsc{Astropy} \citep{Astropy}, in which the Sun is located at $\boldsymbol x_\odot = \{-8.1,\, 0,\, 0.02\}$~kpc and moves with velocity $\boldsymbol v_\odot = \{12.9,\, 245.6,\, 7.8\}$~\kms. The Sgr remnant is centered at $\alpha=283.76^\circ$, $\delta=-30.48^\circ$, its mean proper motion is $\mu_\alpha=-2.7$~\masyr, $\mu_\delta=-1.35$~\masyr, and its line-of-sight velocity is 142~\kms. The distance used in various studies ranges from 24 to 28 kpc. Here we adopt the value 27~kpc, based on the sample of RR Lyrae described in the next section. This is slightly larger than the value 26.5~kpc used in \citetalias{Vasiliev2020}, and it gives somewhat better fits to the stream properties. The Galactocentric position and velocity of the Sgr remnant are thus $\boldsymbol x_\mathrm{Sgr} = \{17.9,\, 2.6,\, -6.6\}$~kpc, $\boldsymbol v_\mathrm{Sgr} = \{239.5,\, -29.6,\, 213.5\}$~\kms.  

\subsection{Stream distance trend with \Gaia RR Lyrae}

The Sgr tails cover a wide range of distances as they wrap around the Milky Way. To correct for the variation in the debris proximity, we rely on the RR Lyrae stars in the stream. We take the union of the two \Gaia DR2 \citep{Brown2018} catalogues of stars classified as RRL, i.e.  the RRLs in the SOS (Specific Object Study, \citealt{Clementini2019}) and the list of RRLs in the general variability table \texttt{vari\_classifier\_result} \citep{Holl2018}, following \citet{Iorio2019,Iorio2020}. Once the duplicates are weeded out, we do not subject the resulting catalogue to heavy cleaning; we apply a single restriction, namely that based on the quality of the photometry using \texttt{phot\_bp\_rp\_excess\_factor}$<3$ \citep[but see also][]{Lindegren2018}. The RRL's extinction in the $G$ band is obtained from the reddening maps of \citet{Schlegel1998} with $A_G=2.27\times E(B-V)$, while the RRL absolute magnitude is assumed to be constant, $M_G=0.64$ (see \citealt{Iorio2019,Iorio2020} for discussion).

The top left panel of Figure~\ref{fig:observational_selection} shows a density plot of the apparent magnitude, $G$, versus the stream longitude, $\Lambda$, for Sgr stream RR Lyrae selected with $|\Beta|<10^{\circ}$ after subtracting the foreground contribution ($20^{\circ}<|\Beta|<40^{\circ}$). The Sgr main body ($\Lambda\approx0^{\circ}$) as well as the leading ($\Lambda>0^{\circ}$) and the trailing ($\Lambda<0^{\circ}$) tails are clearly visible (narrow, dark, zig-zaging band). Additionally, a small portion of the trailing arm wrap is discernible at $\Lambda>140^{\circ}$ and $G>19$. Leaving the distant trailing material aside, we select $360^{\circ}$ worth of the Sgr stream by applying the proper motion cuts described in the next section. Accordingly, the top right panel of Figure~\ref{fig:observational_selection} shows the density of the Sgr stream RR Lyrae after applying cuts in both latitude and proper motion. The distribution is dominated by stream members with minimal foreground contamination. The thick solid black line represents the stream distance ridgeline, obtained by drawing a cubic spline through the debris density peaks.

\subsection{Red Giant stream members}  \label{sec:rgb_sel}

To study the kinematics of the Sgr debris, we concentrate on the Red Giant Branch (RGB) population. These are the brightest stream members and therefore have the lowest proper motion uncertainty. We start by cross-matching the \Gaia DR2 source catalouge with the 2MASS catalogue \citep{Skrutskie2006}. Only objects with Galactic latitude $|b|>5^{\circ}$, extinction-corrected $13<G<18,\, BP-RP>1$ and parallax $\varpi<0.1$ mas as reported by \Gaia were included in the cross-match. The \Gaia--2MASS cross-match yielded $\sim11.5\times10^6$ stars across the whole sky. The likely stream member stars were selected from this sample using cuts in proper motion, \Gaia and 2MASS colours and absolute magnitude. These selection cuts are illustrated in the second and third rows of Figure~\ref{fig:observational_selection}. First, redder stars are picked using $H-K>0.04$, then a combination of cuts in proper motion $(\mu_{\alpha}, \mu_{\delta})$, \Gaia colour--magnitude $(BP-RP, M_G)$ and 2MASS colour $(J-K, H-K)$ are applied (corresponding selection boundaries are shown as thick blue lines). The four central panels of Figure~\ref{fig:observational_selection}, shown in rows two and three, give the behaviour of the likely Sgr stream members (yellow--red contours) overplotted on top of the foreground population (greyscale density). In each of these panels, the Sgr streams members are obtained by requiring $|\Beta|<10^{\circ}$ together with all other criteria mentioned above, excluding the cut involving the properties shown in the panel. The foreground stars were selected by applying the same criteria, except for stars with $20^{\circ}<|\Beta|<40^{\circ}$. We note that the colours and magnitudes are corrected for the effects of dust extinction using the maps of \citet{Schlegel1998}, and the absolute magnitude $M_G$ is computed by subtracting the RR Lyrae distance modulus (as a function of $\Lambda$) from the apparent magnitude $G$. Given the choice of the RRL absolute magnitude described above, the RRL distance modulus is simply $m-M=G_{\rm RRL}(\Lambda)-0.64$.

The distribution of the foreground and stream stars in the space of colour and absolute magnitude is shown in the right panel of the third row of Figure~\ref{fig:observational_selection}. The distance trend as inferred from the RRL behaviour appears to have worked reasonably well, as the stream stars are limited to a narrow RGB sequence (also note a tight Red Clump around $BP-RP\approx 1.1, M_G\approx0.5$) with little overlap with the foreground stars (grey). Stream giants also peel off from the foreground dwarfs when infrared colours are used as revealed in the left panel of the same row. Individually, each of the $BP-RP, M_G$ and $J-K, H-K$ selections reduces the sample by more than a half; applying a combination of these two cuts reduces the sample by nearly $90\%$. However, the most important vetting is that carried out in the proper motion space. The left and right panels in the second row of Figure~\ref{fig:observational_selection} present the proper motion components $\mu_{\alpha}$ and $\mu_{\delta}$ as functions of the stream longitude $\Lambda$. In each of the components the stream sequence is remarkably slender (thanks to the superb quality of the \Gaia astrometry and owing to the bright selection employed). The proper motion sequences show a significant broadening at $-100^{\circ}<\Lambda<0^{\circ}$ caused by the stream's depth along the line of sight. Across the entire range of $\Lambda$, however, the separation between the stream and the foreground is good, notwithstanding some partial overlap at low Galactic latitudes. The proper motion cut alone reduces the sample size drastically, by a factor of $\sim30$. The combination of cuts in proper motion, optical colour, absolute magnitude and near-infrared colours produces a total of $\sim68\,000$ Sgr stream RGB candidates.

The positions of the thus selected RGB candidates are shown in the bottom left panel of Figure~\ref{fig:observational_selection} in the stream-aligned coordinates $\Lambda, \Beta$. While there is some foreground contamination, it is limited to $|b|<10^{\circ}$. Outside of the low Galactic latitudes, our stream sample is rather pure: there are only few stars above and below the trailing tail and a sprinkle of stars underneath the leading one. The tails themselves show plenty of structure. There are hints of a narrow component at $-\Beta=10^{\circ}$ in the trailing tail and several streamlets in the leading tail. The width of the stream (its extent in $\Beta$) is remarkably large: e.g. at $\Lambda\approx -100^{\circ}$ the trailing tail appears to be some $40^{\circ}$ wide. We further reduce foreground contamination by creating a cubic polynomial mask in $(\Lambda, \Beta)$ shown as solid blue line in the bottom left panel of Figure~\ref{fig:observational_selection}. The principal purpose of this rather generous mask is to suppress the interlopers at low Galactic latitudes. The number of candidate stream members is reduced from $\sim68\,000$ to $\sim55\,000$ after applying the $(\Lambda, \Beta)$ mask; of these, $\sim15\,000$ are outside the remnant itself. This represents our final sample of the Sgr RGB stars. Each candidate giant in this sample is assigned a heliocentric distance using the median distance of the 5 nearest RR Lyrae on the sky. We take the median absolute deviation as the uncertainty of this distance estimate.

This dataset with 5D phase-space information can be enhanced with an addition of line-of-sight velocity measurements for some of the stars. We cross-match the sample of the likely Sgr RGB members with the publicly available spectroscopic catalogues. In the end, we have found 4569 line-of-sight velocity measurements, and only six data sources have contributed non-trivial number of matches, these are APOGEE (21\%), LAMOST (13\%), \Gaia RVS (2\%), SDSS (8\%), Simbad (26\%, including the data from \citealt{Frinchaboy2012}) and the catalogue of \citet{Penarrubia2011} (29\%). Heliocentric velocities of the RGB stars within the $(\Lambda, \Beta)$ mask shown in the bottom left panel of Figure~\ref{fig:observational_selection} are plotted in the bottom right panel of the Figure. Most of the stars, more precisely 4465 (of which $\sim1200$ are outside the remnant), follow a well-defined velocity signature of the stream (highlighted with the solid blue lines). This implies only $\sim$3\% contamination of the stream RGB sample. This might be only a lower limit however, as the stars with spectra tend to avoid low Galactic latitudes.

In conclusion, we stress that in essence, our selection of the stream members is not particularly different from those obtained with the use of the \Gaia DR2 and reported in the literature recently \citep[e.g.][]{Antoja2020, Ibata2020}. The main distinction is that we have limited our sample to the brightest members of the stream, the RGB stars -- this allows us to fine-tune the selection and filter out most of the intervening dwarf stars. The second important (and unique) feature of our approach is using the well-determined stream's distance variation to our advantage.  To be more quantitative in our comparison, we point out that the sample contamination reported here is some $4-5$ times smaller than that reported in \citet{Ibata2020} who used a very similar idea of employing the spectroscopy to test for the presence of interlopers (they find 14\% contamination for stars with $G<17$ and very similar sky coverage). The stream member selection presented in \citet{Antoja2020} relies on the proper motion only, therefore, naturally, we would expect a larger fraction of false positives for such a sample. Additionally, as the stream is detected automatically at each location on the sky, some of their detections may suffer a kinematic bias due to the overpowering foreground presence, as indeed mentioned in \citet{Antoja2020}.

\subsection{Velocity misalignment}  \label{sec:misalignment}

\begin{figure}
\includegraphics{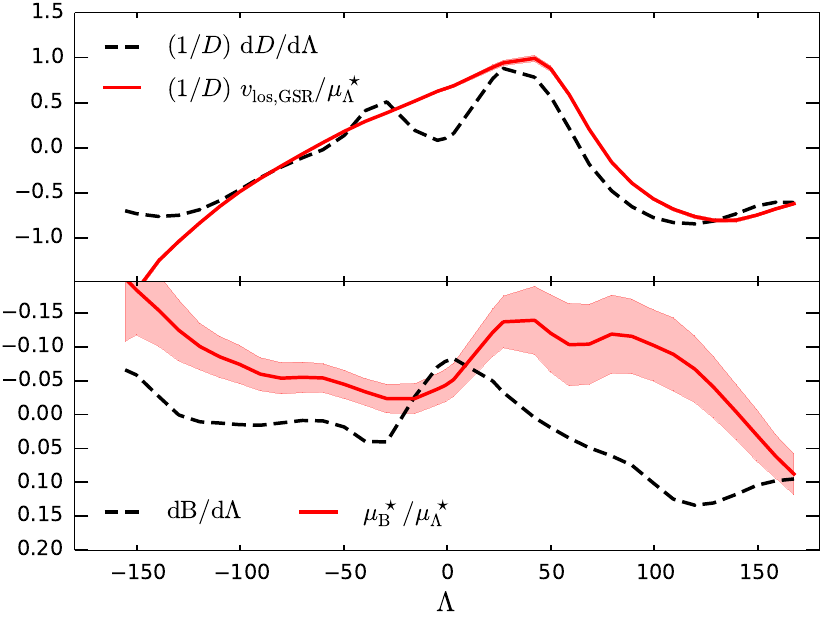}
\caption{Velocity misaligment in the Sgr stream in two projections: longitude vs.\ heliocentric distance (top panel) and longitude vs.\ latitude (bottom panel). Dashed black lines show the slope of the stream track in these coordinates, while solid red lines show the ratio of corresponding velocity components (reflex-corrected). Systematic uncertainties arising from varying the Solar velocity and distance modulus are depicted by red shaded regions (they are negligible in the top panel, being of order a few percent only). See Section~\ref{sec:misalignment} for further discussion.
}  \label{fig:misalignment}
\end{figure}

Having determined proper motion and distance trends along the entire stream, we now analyse whether the mean velocity is aligned with the stream, following the approach of \citet{Erkal2019}. For this exercise, we need to correct the observed proper motions for the Solar reflex motion, which can be done since we know the (mean) heliocentric distance at each point. If the stars indeed move along the stream, the ratio of reflex-corrected proper motion components $\mu_\Beta^\star / \mu_\Lambda^\star$ at any point should equal the gradient of the stream track on the sky $\mathrm{d}\Beta/\mathrm{d}\Lambda$. Similarly, the ratio of the reflex-corrected line-of-sight velocity $v_\mathrm{los,GSR}$ to $\mu_\Lambda^\star$ should equal the gradient of the distance along the stream $\mathrm{d}D/\mathrm{d}\Lambda$. 

Figure~\ref{fig:misalignment} shows these quantities plotted as functions of the stream longitude $\Lambda$. While the two curves in the top panel follow each other rather closely (except the region around $\Lambda=0^{\circ}$, where the above arguments do not hold because of internal motions in the remnant), the bottom panel shows a clear misalignment between the stream track and proper motions, especially in the leading arm ($\Lambda>0$). Red shaded region shows the uncertainties in the ratio of proper motions arising from the Poisson noise and from a systematic variation of the Solar velocity (by 5~\kms) and the heliocentric distance to the stream (by 2.5\%, i.e., 0.05~mag calibration error in the RR Lyrae magnitudes). The two sources of systematic uncertainty have a roughly equal contribution to the error budget, which dominates the Poisson noise thanks to the large number of stars in the stream. We caution that the misalignment in the trailing arm is rather sensitive to the sample of stars used in the analysis. The large relative thickness of the stream along the line of sight broadens the distribution of proper motions and makes it asymmetric, with a long tail of stars at small distances and correspondingly high proper motions. Since we use mean distance to the stream for reflex correction, stars in these tails would be undercorrected and may bias the mean $\mu_\Beta^\star$ quite significantly. Therefore, the systematic errors in $\mu_\Beta^\star/\mu_\Lambda^\star$ may be underestimated in the trailing arm, but the leading arm, being much further away, is less prone to these distortions, and the apparent misalignment must be real and highly significant.

Such a misalignment has been previously found for the Orphan stream \citep{Koposov2019,Erkal2019}, and is also detected for several streams in the Galactic southern hemisphere \citep{Shipp2019,Li2020}. Interestingly, no misalignment was found for the GD-1 stream \citep{deBoer2020}, which is in the northern hemisphere. While the stream track may precess on the sky in a non-spherical Galactic potential \citep[e.g.][]{Erkal2016}, a close alignment of the velocity with the stream track is expected in any static potential. Therefore, the observed misalignment is a smoking-gun signature of a time-dependent perturbation to the Sgr stream, and the most likely (though not the only possible) explanation pursued in this paper is the influence of the LMC -- the same scenario that was used to explain the Orphan stream by \citet{Erkal2019}. 

We note that the expected alignment between the Sgr stream track and its reflex-corrected proper motions was used by \citet{Hayes2018} to constrain the Solar velocity (since the amount of reflex correction directly depends on it). Although the value they obtained is quite close to other recent independent measurements, the assumption behind this analysis is invalidated by a time-dependent potential.

\section{Modelling approach}   \label{sec:method}

\subsection{Previous work}

Previous studies employed a variety of approaches for fitting the stream properties. The simplest one is to compute the orbit of a test particle in the given potential \citep[see e.g.][]{Koposov2010}. However, while the tidal streams from low-mass stellar systems such as globular clusters follow their orbits reasonably closely (although see \citealt{Eyre2011}, \citealt{Sanders2013}), this approximation is entirely inadequate for such a massive progenitor as Sgr (e.g., \citealt{Choi2007}).

The next level of complexity is achieved by methods that take into account the offset of stripped debris from the orbit of the progenitor. These algorithms either strive to reproduce the mean track of the stream \citep[sometimes called a {\it streakline} for its characteristic appearance, see][]{Varghese2011,Lane2012,Kupper2012,Bonaca2014} or the individual stream members \citep[a.k.a. {\it modified Lagrange Cloud Stripping} or mLCS, see][]{Gibbons2014}.  Subsequently, \citet{Fardal2015} compared several stream generating methods and demonstrated that mLCS provided a fast and realistic description for streams similar to Sgr. However, they argued that the technique could be refined further and accordingly introduced their improvements in the method dubbed {\it particle spray}. Recently, an upgraded version of mLCS has been used successfully to model a wide variety of Galactic streams \citep[][]{Bowden2015,Erkal2017,Erkal2018,Erkal2019}. mLCS is a constrained $N$-body method, which places the escaping particles at some distance from the progenitor centre (comparable to its tidal radius) and integrates their orbits in the potential of the host galaxy, while also taking the satellite's gravity into account. It is fast, and although in its basic version it assumes simple models for the progenitor (e.g., a spherical system with some mass, radius and a fixed density profile), the evolution of satellite properties can be easily included. What mLCS can not currently be asked to do is to realistically reproduce the properties of the remnant. Therefore, here, faced with the wealth of the currently available information on the stream and the remnant, we choose to start with a fast and approximate mLCS-like method and move to the next logical step, involving full $N$-body simulations.

Full-fledged $N$-body simulations have been used in many studies of the Sgr stream. In some variants (e.g., \citealt{Helmi2004}, \citetalias{Law2010a}, \citealt{Penarrubia2010}), the particles belonging to the Sgr galaxy (as well as those that have been stripped) feel their own gravitational force (using a tree-code or a basis-set expansion as the Poisson solver) and a static, external potential of the Milky Way from an analytically specified mass distribution. In this case, the initial conditions for the orbit are usually obtained by integrating the motion of a test particle in the Milky Way potential back in time, but the actual orbit deviates slightly from this test-particle orbit due to the gravitational torques from the stripped particles (``dynamical self-friction'', e.g., \citealt{Miller2020}). However, the ``classical'' dynamical friction from the Milky Way halo is not accounted for, which is believed to be an acceptable approximation at the last stages of Sgr evolution. In other cases (e.g., \citealt{Lokas2010}, \citealt{Dierickx2017}), the Milky Way is also represented as a live $N$-body system, automatically providing the correct amount of dynamical friction. However, the final position and velocity is even harder to control in this case, which means that the models are not intended to represent the actual Sgr stream in detail (for instance, the \citealt{Dierickx2017} model missed the true position by several kpc, the velocity -- by tens of \kms, and the orbital plane -- by some $30^\circ$).

The dynamical influence of the LMC on the Sgr stream was considered only in a couple of papers. \citet{VeraCiro2013} used the test-particle approximation for the orbit of Sgr in the static Milky Way potential plus a moving LMC following a test-particle orbit in the same Milky Way potential. The leading and trailing arms of the stream were constructed by following the Sgr orbit from its present-day position backward and forward in time in this evolving potential. In addition to the inadequacy of the test-particle approach to represent the actual stream, this also effectively uses the future orbit of the LMC to compute the trajectory of the leading arm, and ignores the reflex motion of the Milky Way. Nevertheless, they were able to demonstrate that the inclusion of the LMC has an important effect on the Sgr stream and on the inferred shape of the Milky Way potential.

Subsequently, \citet{Gomez2015} explored the more realistic situation when the Milky Way is allowed to move in response to the gravitational pull of the LMC, and found that it can acquire a reflex velocity of several tens \kms in a short time around the pericentre passage of the LMC, which occurred just recently. Such a significant acceleration of the reference frame inevitably displaces the Sgr stream, and they demonstrated that its inferred orbital parameters (eccentricity, precession rate) change very substantially. They represented both the Milky Way and Sgr as live $N$-body systems, and the LMC -- as a single softened moving particle. Although they did not attempt to match in detail the current position and velocity of the Sgr remnant, their simulations demonstrated that the inclusion of the LMC may obviate the need for a triaxial Milky Way potential to explain the Sgr stream properties. 

Finally, \citet{Cunningham2020} present a fit to the Sgr stream in the presence of the LMC using the mLCS technique as implemented in \citet{Erkal2019}. This approach treats the Milky Way and LMC each as particles sourcing their respective potentials and thus naturally allows for the reflex motion of the Milky Way. The dynamical friction from the Milky Way on both Sgr and the LMC is included using the results of \citet{Jethwa2016}. The Milky Way halo is represented by the triaxial generalization of the NFW profile from \citet{Bowden2013}, which allows for a different inner and outer halo flattening. They found that an LMC mass of $\sim2\times10^{11} M_\odot$ was needed to reproduce the observational constraints on the Sgr stream from \citet{Belokurov2014}. We note while their approach is broadly similar to the restricted $N$-body approach described in Section~\ref{sec:fast_stream_generation}, the present study uses different methods and is completely independent.

\subsection{$N$-body simulations}  \label{sec:Nbody_simulations}

Following the above arguments, we choose to use fully self-consistent $N$-body simulations to follow the evolution of the Sgr remnant and the stream as our primary tool. However, we rely on certain approximations to represent the effects of the LMC, reflex motion, and dynamical friction.

\begin{figure}
\includegraphics{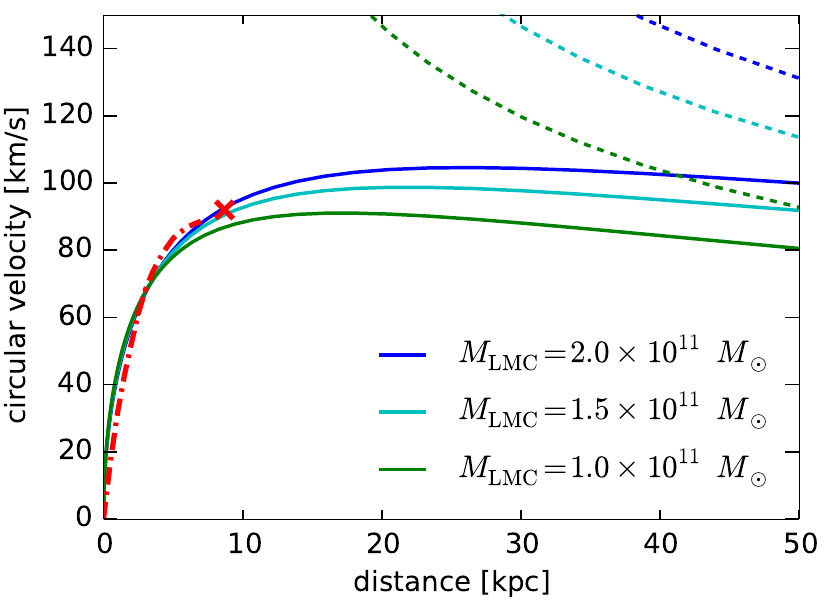}
\caption{Circular velocity curves of LMC models with different masses (solid lines, from top to bottom: $\{2, 1.5, 1\}\times10^{11}\,M_\odot$). 
Note that the mass enclosed within a few tens of kpc is considerably smaller than the total mass, shown by short-dashed lines corresponding to Keplerian rotation curves. Red dot-dashed line shows the observational constraints from \citet{Vasiliev2018}, and the red cross shows the measurement of \citet{vdMarel2014}.
} \label{fig:vcirc_lmc}
\end{figure}

First, for a given choice of the Milky Way potential and the LMC mass, we run a simulation of a live Milky Way plus the LMC with the $N$-body code \textsc{Gyrfalcon} \citep{Dehnen2000}, which is included in the \textsc{Nemo} stellar-dynamical framework \citep{Teuben1995}. The parameters of the Milky Way models are described in Section~\ref{sec:MW_model}; the disc and bulge together are represented by $10^6$ particles, and the halo -- by $4\times10^6$ particles. 
The LMC is modelled as a truncated spherical NFW model with a density profile
\begin{equation}  \label{eq:lmc_density}
\rho_\mathrm{LMC} \propto r^{-1}\;(1+r/r_\mathrm{scale})^{-2}\;\exp\big[-(r/r_\mathrm{trunc})^2\big] .
\end{equation}
We set $r_\mathrm{trunc}=10\,r_\mathrm{scale}$ and choose the scale radius so that the circular velocity $v_\mathrm{circ}\equiv \sqrt{r\,\partial\Phi/\partial r}$ in the inner part of the galaxy satisfies observational constraints (Figure~\ref{fig:vcirc_lmc}): $r_\mathrm{scale}=8.5\,\mbox{kpc}\times(M_\mathrm{LMC}/10^{11}\,M_\odot)^{0.6}$. We sample the LMC model with $10^6$ particles.

We need to reproduce the present-day position and velocity of the LMC, but there is some ambiguity in defining the LMC centre \citep[e.g.,][]{vdMarel2014,Helmi2018}, and different reference points imply different values of the mean proper motion. We experimented with different choices of these values, and found that they have relatively little impact on the appearance of the Sgr stream (comparable to the variations between Milky Way potentials within the range explored in Section~\ref{sec:potential}). Following \citet{Vasiliev2018}, we adopt the photometric centre of the bar $\alpha_\mathrm{LMC}=81^\circ$, $\delta_\mathrm{LMC}=-69.75^\circ$ as our reference point, which corresponds to the proper motions $\mu_{\alpha,\mathrm{LMC}}=1.8$~\masyr, $\mu_{\delta,\mathrm{LMC}}=0.35$~\masyr. We take the distance to the LMC equal to 50~kpc \citep{Freedman2001} and the line-of-sight velocity equal to 260~\kms \citep{vdMarel2014}; this translates to the Galactocentric position and velocity $\boldsymbol x_\mathrm{LMC,0} = \{-0.6,\, -41.3,\, -27.1\}$~kpc, $\boldsymbol v_\mathrm{LMC,0} = \{-63.9,\, -213.8,\, 206.6\}$~\kms.

We iteratively adjust the initial position and velocity of the LMC at $T_\mathrm{init}=-3$~Gyr so that its final position and velocity relative to the moving Milky Way reference frame match the observational constraints within 0.2~kpc and 1~\kms. To achieve this, we follow the method detailed in the Appendix of \citetalias{Vasiliev2020}. In brief, for each iteration we run 6 simulations with slightly perturbed initial conditions, and then use the Gauss--Newton method for solving a nonlinear system of equations (determining the initial conditions that produce the desired final state), which converges in $3-4$ iterations.

Of course, the central parts of the Milky Way move in response to the gravitational pull of the LMC;
the net velocity reaches several tens of \kms during the last Gyr. We find it is more convenient to work in the reference frame centered on the Milky Way, so we construct a smooth approximation for the trajectory of the Milky Way centre, and differentiate it twice to obtain the time-dependent but spatially uniform acceleration field associated with this non-inertial frame.

We then turn to simulating the Sgr system over the same interval of time (3 Gyr), corresponding roughly to 2.5 orbital periods for our most suitable potential parameters. We again employ the \textsc{Gyrfalcon} $N$-body code, which has an option to include external forces. In \citetalias{Vasiliev2020}, we used just the static Milky Way potential for this purpose (represented by the \textsc{Agama} plugin for \textsc{Nemo}). We now add further components to this external force and make it time-dependent. In the simplest case, the LMC is represented by a fixed analytic potential corresponding to the initial density profile (\ref{eq:lmc_density}), and moves along its pre-recorded trajectory translated into the Milky Way-centered frame. The acceleration associated with this non-inertial frame is also applied to all particles in the simulation. However, such a massive and spatially extended LMC does experience tidal distortion during its encounter with the Milky Way. The Galactic halo is also perturbed by the LMC in two ways: a ``local'' wake is formed behind the LMC and is responsible for the dynamical friction on the latter, and a large-scale ``global'' wake results from the differential motion of the inner part of the Galaxy with respect to the outer halo (see figure~1 in \citealt{GaravitoCamargo2020}). We may account for these distortions by constructing a series of basis-set approximations to the time-dependent LMC and Milky Way halo density profiles, as detailed in \citet{Sanders2020}. We use the Multipole potential from \textsc{Agama} with the order of angular expansion $l_\mathrm{max}=4$. These distortions have a minor but noticeable effect on the Sgr orbit, as discussed in Section~\ref{sec:with_LMC}. This approach, with some variations, has been used for resimulating streams in pre-recorded host galaxy potentials \citep[e.g.][]{Lowing2011,Ngan2015,Dai2018}.

We also account approximately for the dynamical friction acting on the Sgr itself, using the following kludge. At each block timestep of the simulation, we estimate the position $\boldsymbol x_\mathrm{Sgr}$ and velocity $\boldsymbol v_\mathrm{Sgr}$  of the Sgr centre, taking the median of the values for particles within 5 kpc from the centre position extrapolated from the previous timestep, and then iteratively refining it. We use the median instead of the mass-weighted mean to avoid biases caused by spatially overlapping portions of the stream from earlier stripping episodes, which are always subdominant in the number of particles, but may have widely different velocities. Then we take the total mass of particles within 5 kpc as the estimate of the current remnant mass $M_\mathrm{Sgr}$, and apply the Chandrasekhar prescription for the dynamical friction-induced acceleration:
\begin{eqnarray}  \label{eq:dynfric}
\boldsymbol a_\mathrm{DF} &\!\!\!=\!\!\!& -4\pi\,\rho_\mathrm{MW}(\boldsymbol x_\mathrm{Sgr}) \,
\ln\Lambda\; G\, M_\mathrm{Sgr}\, \boldsymbol v_\mathrm{Sgr} \big/ |v_\mathrm{Sgr}|^3 \times {} \nonumber \\
&& \big[ \mathrm{erf}(X) - 2\pi^{-1/2}\,X\,\exp(-X^2) \big] ,\\
X &\!\!\!\equiv\!\!\!&  |v_\mathrm{Sgr}| \big/ \sqrt{2}\,\sigma_\mathrm{MW}, \nonumber
\end{eqnarray}
where $\rho_\mathrm{MW}$ and $\sigma_\mathrm{MW}$ are the Milky Way halo density and velocity dispersion (we take a fixed value 120~\kms for the latter for simplicity, noting that the result is rather insensitive to the precise value), and $\ln\Lambda$ is the Coulomb logarithm. We choose the value of this parameter by running a calibration suite of simulations with a live Milky Way, and find that setting $\ln\Lambda=3$ matches the orbital decay rate and mass loss fairly well. We note that even in absence of explicitly added drag force, the Sgr orbit still decays due to the ``self-friction'' from the previously stripped mass, which is, however, a few times weaker than the Milky Way-induced friction. One advantage of our implementation compared to the similar analytic dynamical friction prescriptions used in mLCS-like methods is that we track the remnant mass realistically in the actual simulation.
The moving Sgr also causes the reflex motion of the Milky Way, however, its amplitude is much smaller than that from the LMC (at the level 1~kpc and $1-2$~\kms), so we ignore these offsets in the analysis.

To model the properties of the Sgr stream adequately, it is essential to match the present-day properties of the Sgr remnant and its current position and velocity. We use the same iterative technique to find the orbital initial conditions, modified to account for the time-dependent potential. In \citetalias{Vasiliev2020}, we located the point on each orbit that is closest to the desired final state, and used the time offset between this point and the desired final time $T=0$ to shift the initial conditions by integrating a test-particle orbit for the same interval of time. This is no longer possible if the potential is time-dependent: an orbit started from the same initial point at a later time will not follow the same trajectory. Hence we explicitly add another ``delayed'' orbit with the same initial conditions to the list of perturbed orbits considered in each iteration, and estimate the complete 6d Jacobian matrix that relates the initial state to the final state. A few iterations are needed to match the final position and velocity to an accuracy of 0.1~kpc and 1~\kms.

As shown in \citetalias{Vasiliev2020}, the observed kinematic features of the Sgr remnant (the spatial variation of the mean proper motion, line-of-sight velocity, and dispersions of these quantities) place strong constraints on its present-day structure. The total mass needs to be in the range $(3-5)\times10^8\,M_\odot$ in order to match the velocity dispersions; the stellar mass is estimated photometrically to be around $10^8\,M_\odot$, and the remnant must have an elongated shape with the angle between the long axis and the velocity vector around $45^\circ$, extending up to $\sim 5$~kpc from its centre. We verify that these conclusions remain valid in our more complicated scenario with time-dependent perturbations from the LMC. 

Our $N$-body approach has several advantages:
\begin{itemize}
\item Follows the internal dynamics of the Sgr galaxy self-consistently (unlike mLCS-like techniques), hence we may constrain the model parameters by the observed properties of the Sgr remnant, as in \citetalias{Vasiliev2020}.
\item Takes into account (approximately) the dynamical friction of the Sgr orbiting the Milky Way, and follows the mass loss rate realistically.
\item Includes the following effects arising due to the presence of the LMC: its own gravitational field, the reflex motion of the Milky Way, and optionally the gravitational wake in the Milky Way halo and the evolving LMC shape.
\item Does not assume that the remnant follows a test-particle orbit in the given potential, but constrains the orbital initial conditions by the present-day position and velocity of the remnant to high accuracy.
\end{itemize}
The main disadvantage, of course, is the computational cost: for each choice of Milky Way and LMC parameters, we need to run a full $N$-body simulation of the LMC orbit, and then for each choice of the Sgr parameters, run another $N$-body simulation. In fact, since we fit the present-day position and velocity of both the LMC and the Sgr to within a fraction of a kpc and a \kms, we need to perform several iterations of simulations, adjusting the orbital initial conditions (and in the case of the Sgr, also its structural properties) to match the current state. For this reason, we cannot perform a comprehensive exploration of the parameter space, but rather study the effects of changing one parameter at a time with a few dozen models.

\subsection{Stream generation by restricted $N$-body simulations}  \label{sec:fast_stream_generation}

Due to the high cost of the primary $N$-body simulation method, we complement it with another scheme, which is more approximate but much faster. It has a lot in common with the stream generation approaches used by \citet{Gibbons2014} and \citet{Fardal2019}. Namely, we follow the evolution of an ensemble of test particles in the fixed gravitational potential of the Milky Way plus the moving potential of the Sgr remnant.

First, for any choice of the Milky Way potential parameters and the LMC mass, we compute the approximate trajectory of the LMC and the corresponding reflex motion of the Milky Way, by integrating a coupled system of differential equations
\begin{equation}  \label{eq:lmc_orbit}
\begin{aligned}
\dot{\boldsymbol{x}}_\mathrm{MW} &= \boldsymbol{v}_\mathrm{MW}, \\
\dot{\boldsymbol{v}}_\mathrm{MW} &= -\nabla\Phi_\mathrm{LMC}(\boldsymbol{x}_\mathrm{MW}-\boldsymbol{x}_\mathrm{LMC}), \\
\dot{\boldsymbol{x}}_\mathrm{LMC} &= \boldsymbol{v}_\mathrm{LMC}, \\
\dot{\boldsymbol{v}}_\mathrm{LMC} &= -\nabla\Phi_\mathrm{MW}(\boldsymbol{x}_\mathrm{LMC}-\boldsymbol{x}_\mathrm{MW}) + \boldsymbol{a}_\mathrm{DF}, 
\end{aligned}
\end{equation}
backward in time starting from the current LMC position and velocity. Here $\boldsymbol{a}_\mathrm{DF}$ is the dynamical friction acceleration as defined by Equation~\ref{eq:dynfric}, but for the LMC rather than Sgr. We then transform the trajectory of the LMC to the non-inertial reference frame centered on the Milky Way, and initialize the corresponding time-dependent spatially-uniform acceleration of this reference frame.

The trajectory of the Sgr remnant is computed by integrating the orbit backward in time, taking into account dynamical friction force (Eqn.~\ref{eq:dynfric}) with a time-dependent remnant mass parametrized by a simple expression. By analysing the actual $N$-body simulations, we conclude that the mass drops precipitously during $\sim 0.3$~Gyr after each pericentre passage, and then stays constant until the next one. We thus approximate the remnant mass by a piecewise-linear function with two equal drops over the preceding pericentre passages (excluding the most recent one, which occurred just $\sim 0.05$~Gyr ago), and represent its potential by a Plummer sphere with the given mass and a scale radius of $\sim1.5$~kpc (this value defines the depth of the potential well, and changes rather mildly during the run).

The particles are initially taken from the same two-component progenitor model used in the full $N$-body simulations, and are evolved in the time-dependent analytic potential composed of the Milky Way, moving Sgr, moving LMC, and the spatially uniform acceleration of the non-inertial reference frame. Particles staying close to the progenitor centre move with it, while those further out may escape and form the stream; in all cases, the particles feel the gravity of the progenitor, and are automatically placed on realistic escape orbits. The escape rate and hence the distribution of particles along the stream may be somewhat distorted in comparison with the actual $N$-body simulation, but the overall configuration of the stream is reproduced fairly well. Figure~\ref{fig:stream_compare_restricted} in the Appendix shows the comparison of observed stream properties in the two approaches, for a Milky Way + LMC model discussed in Section~\ref{sec:with_LMC}.

This approach is fast enough (taking $\mathcal O(10^2)$ CPU seconds for $10^5$ particles, with a near-perfect parallelization speedup) to make possible a Markov Chain Monte Carlo (MCMC) exploration of the parameter space, which we perform with the code \textsc{emcee} \citep{ForemanMackey2013}. For the score function, we use the differences between the model-predicted and observed quantities in $\sim 30$ bins in $\Lambda$: heliocentric distance, line-of-sight velocity, sky-plane track (mean $\Beta$), and the ratio of reflex-corrected mean proper motions $\mu_\Beta^\star/\mu_\Lambda^\star$. We assume a $10\%$ distance uncertainty, a $12$~\kms velocity uncertainty, and a $3^\circ$ uncertainty in $\Beta$ (the actual measurements have considerably smaller error bars, but we allow for some model mismatch due to unaccounted factors).

\subsection{Milky Way models}  \label{sec:MW_model}

The Milky Way model consists of a spherical bulge, an exponential disc, and a dark halo. The bulge density profile is 
\begin{equation}
\rho_\mathrm{b} \propto \big(1+r/r_\mathrm{b}\big)^{-\Gamma}\, \exp\big[-(r/u_\mathrm{b})^2],
\end{equation}
with $r_\mathrm{b} = 0.2$~kpc, $u_\mathrm{b} = 1.8$~kpc, $\Gamma=1.8$, and total mass $1.2\times10^{10}\,M_\odot$. The disc follows the exponential/isothermal density profile:
\begin{equation}
\rho_\mathrm{d} \propto \exp\big[-R/R_\mathrm{d}\big]\, \mathrm{sech}^2\big[z/(2h)\big],
\end{equation}
with a total mass $5\times10^{10}\,M_\odot$, a scale radius $R_\mathrm{d} = 3$~kpc, and a scale height $h_\mathrm{d} = 0.4$~kpc. We keep the parameters of the stellar distribution fixed and close to the commonly used values suggested by \citet{McMillan2017}.
By contrast, the halo density profile is very flexible:
\begin{equation}  \label{eq:halo_density}
\begin{aligned}
\rho_\mathrm{h} &\propto (s/r_\mathrm{h})^{-\gamma}\, 
\big[ 1 + (s/r_\mathrm{h})^\alpha \big]^{(\gamma-\beta)/\alpha}\, \exp\big[-(s/u_\mathrm{h})^\eta], \\
s &\equiv (pq)^{1/3}\,\sqrt{X^2 + (Y/p)^2 + (Z/q)^2}.
\end{aligned}
\end{equation}
Here $s$ is the ``sphericalized radius'', which is equivalent to the usual spherical radius $r$ when the axis ratios $p=q=1$, and approximately corresponds to $r$ when averaged over angles even for $p,q\ne 1$. The sphericalized density follows the \citet{Zhao1996} double-power-law profile, adorned with an exponential cutoff around a radius $u_\mathrm{h}$ with an adjustable steepness $\eta$. The Sgr stream provides little constraints on the halo density beyond $\sim100$~kpc, so the cutoff is simply a practical measure to ensure a finite halo mass, and in most runs we fix fiducial values $u_\mathrm{h}=200$~kpc and $\eta=2$. Thus the trajectory of the LMC at early times may not be fully realistic, but its recent interaction with the Milky Way occurs at a distance $\lesssim 50$~kpc and is insensitive to the details of the outer halo. However, to explore the constraints on the total Milky Way mass, we allow the cutoff parameters to be varied. The normalization of the halo density profile is chosen in such a way as to produce a circular velocity at the Solar radius $v_\mathrm{circ}(R) \equiv \sqrt{R\,\partial\Phi/\partial R}$ close to $235$~\kms -- a value well constrained by many recent studies, and matching \citet{McMillan2017}.

A novel aspect of our study is that we allow the axis ratios $p,q$ and the orientation of the principal axes $X,Y,Z$ to vary with radius, as the preliminary tests suggested that constant-shape haloes do not fit the stream well. The coordinates $\boldsymbol X\equiv \{X,Y,Z\}$ are related to the usual Galactocentric cartesian coordinates $\boldsymbol x\equiv \{x,y,z\}$ by a rotation matrix parametrized by three Euler angles $\alpha_q$, $\beta_q$, $\gamma_q$ (subscripted to distinguish them from the parameters of the \citet{Zhao1996} density profile):
\begin{equation}  \label{eq:halo_orientation}
\begin{aligned}
\boldsymbol X &= \mathsf R \boldsymbol x, \\
\mathsf R &\equiv \left(\!\! \begin{array}{ccc}  \phantom{-}
 c_\alpha c_\gamma - s_\alpha c_\beta s_\gamma\;\; & \phantom{-}
 s_\alpha c_\gamma + c_\alpha c_\beta s_\gamma\;\; &
 s_\beta  s_\gamma \\
-c_\alpha s_\gamma - s_\alpha c_\beta c_\gamma\;\; &
-s_\alpha s_\gamma + c_\alpha c_\beta c_\gamma\;\; &
 s_\beta  c_\gamma \\
 s_\alpha s_\beta &
-c_\alpha s_\beta &
 c_\beta 
\end{array} \!\!\right),
\end{aligned}
\end{equation}
where $c_\circ, s_\circ$ stand for $\cos\circ$, $\sin\circ$.
The axis ratios $p, q$ and the angle $\beta_q$ respectively change from 1, $q_\mathrm{in}$, 0 in the inner part to $p_\mathrm{out}$, $q_\mathrm{out}$ and $\beta_\mathrm{out}$ in the outer part according to the following changeover function:
\begin{equation}
\circ(r) = \big[ \circ_\mathrm{in} + \circ_\mathrm{out}\,(r/r_q)^2 \big] \big/ \big[ 1 + (r/r_q)^2 \big].
\end{equation}
In other words, the inner part (at $r\ll r_q$) is axisymmetric and aligned with the disc plane, satisfying the stability arguments of \citet{Debattista2013}, while the outer part is triaxial, its $Z$ axis is inclined by angle $\beta_\mathrm{out}$ w.r.t.\ the $z$ axis, and the angles $\alpha_q$ ($\gamma_q$) measure the rotation between the $x$ ($X$) axes and the line of intersection of $xy$ and $XY$ planes (see the scheme on Figure~\ref{fig:cornerplot_shape} later in the paper). The angles $\alpha_q, \gamma_q$ do not change with radius, but they are irrelevant for the inner halo since it is axisymmetric. If $\beta_\mathrm{out}\ne 0$, the resulting profile is twisted and non-triaxial, meaning that it cannot be represented by an equilibrium model needed for live $N$-body simulations, but can be used with the restricted $N$-body simulation approach.

In the fully general case, the halo profile has 13 free parameters (or 11 if we fix $u_\mathrm{h}$ and $\eta$ to their fiducial values): inner density slope $\gamma$, outer slope $\beta$, transition steepness $\alpha$, scale radius $r_\mathrm{h}$, outer cutoff radius $u_\mathrm{h}$, cutoff steepness $\eta$, inner and outer axis ratios $q_\mathrm{in},\,q_\mathrm{out},p_\mathrm{out}$, Euler angles $\beta_\mathrm{out}, \alpha_q, \gamma_q$, and the shape transition radius $r_q$. Non-twisted triaxial haloes have 9 parameters ($\beta_\mathrm{out}$ has to be zero, and we may set $\gamma_q=0$ without loss of generality, since only the sum $\alpha_q+\gamma_q$ determines the angle between $x$ and $X$ axes). If we further restrict the density to be axisymmetric (but with a varying axis ratio $q$), this removes two more parameters ($p_\mathrm{out}=1$, $\alpha_q$ irrelevant).
Although we do not have explicit expressions for the potential generated by this complicated density profile, it can be easily evaluated with the general-purpose multipole potential solver; the advantage of specifying the flattening in the density rather than in the potential is that we can ensure physical validity of the profile (it never goes to negative values).

The Milky Way models for a live $N$-body simulation (with a non-twisted halo) are constructed with the \citet{Schwarzschild1979} orbit-superposition approach, as implemented in the \textsc{Agama} framework \citep{Vasiliev2019}. Both stellar and dark halo components are built with $10^5$ orbits each, and the halo velocity distribution is constrained to be approximately isotropic. As demonstrated in \citet{Vasiliev2015}, the Schwarzschild method produces $N$-body models that are very close to equilibrium, in contrast to many alternative approaches, which require an initial period of ``relaxation'' (e.g., Section~3.2.1 in \citealt{GaravitoCamargo2019}). Our models remain stable when evolved in isolation for several Gyr. 

\subsection{Sgr models}  \label{sec:Sgr_model}

We use the same approach as in \citetalias{Vasiliev2020} to construct two-component models of the Sgr progenitor. The stellar density follows a King profile (exactly in the case of spherical models, and approximately for flattened and rotating ones) with the King parameter $W_0=4$, and is embedded in a more extended spherical dark halo. We explored several variants of halo profiles, and found that cored profiles are preferred. The primary reason is that the remnant has a relatively low stellar velocity dispersion at present, but still must be massive and extended enough to avoid being ripped apart earlier. As shown in that paper, models with cuspy haloes are unable to satisfy both constraints simultaneously.

Our fiducial model has an initial stellar mass of $2\times10^8\,M_\odot$ and a halo mass $18\times$ larger; by the present time, approximately half of the stars are stripped, while the total mass of the remnant is $\sim(3-4)\times10^8\,M_\odot$. We have also considered other models with the same stellar mass but lower dark halo masses, and they produced similarly looking streams.  There is no doubt that the initial mass of the Sgr galaxy at infall was still much larger than in our setup (cosmological abundance matching suggest values of a few$\times10^{10}\,M_\odot$, see e.g.\ \citealt{Jiang2000,NiedersteOstholt2010,Gibbons2017,Dierickx2017}). On the other hand, models with a much more gradual mass loss, such as \citetalias{Law2010a}, in which the Sgr mass only decreased from $6.4\times10^8\,M_\odot$ to $2.5\times10^8\,M_\odot$ in the last 8 Gyr, are unable to match the present-day state of the remnant, as shown by \citetalias{Vasiliev2020} (Figure A2). Since we simulate only the last 2.5 orbital periods of its evolution, we truncated the dwarf's halo around the tidal radius at the apocentre of its orbit ($\sim10$~kpc). The stellar and dark matter components of the Sgr galaxy are sampled with $2\times10^5$ particles each. For the latter, we employ a mass-refinement scheme, where particles have different weight depending on their energy, so that more massive but less numerous particles in the outer layers are stripped earlier.

\begin{figure*}
\includegraphics{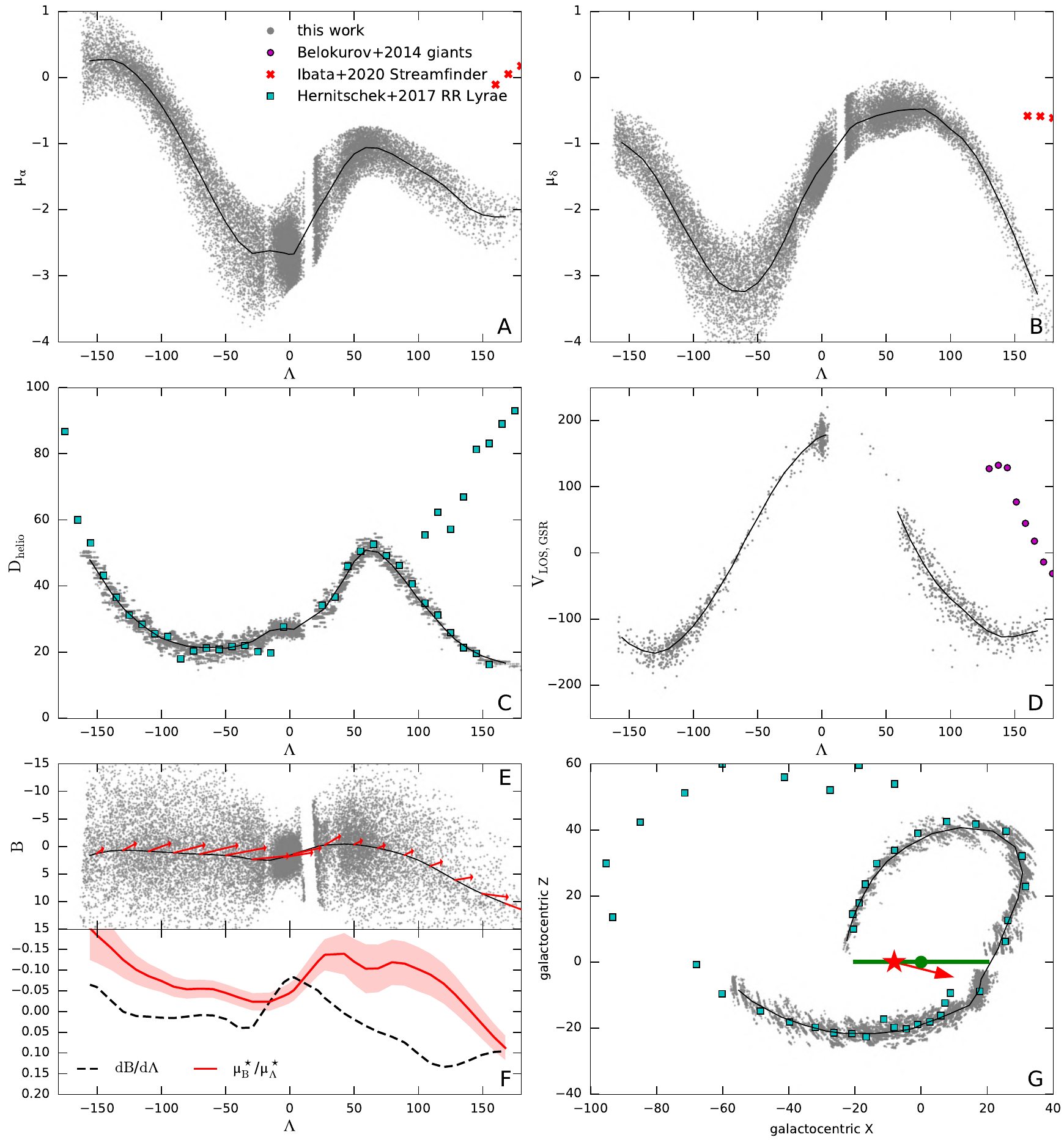}
\caption{Observed properties of the Sgr stream, based on the selection of red giants with distances assigned from RR Lyrae, as described in Section~\ref{sec:data} (grey dots). Panels A and B shows two proper motion components $\mu_\alpha, \mu_\delta$; panel C -- heliocentric distance; panel D -- galactic standard-of-rest line-of-sight velocity; panel E -- sky position; panel G -- Galactocentric side-on view. We augment this dataset with auxiliary measurements from other studies, mainly near the trailing arm apocentre, which is missing in our sample. Red crosses in panels A and B additionally show a few datapoints from \citet{Ibata2020}. Cyan squares in panels C and G show the binned RR Lyrae distance estimates from \citet{Hernitschek2017}. Magenta circles in panel D show line-of-sight velocities from \citet{Belokurov2014}. Red arrows in Panel E plot the reflex-corrected proper motion components in the Sgr coordinates $\Lambda, \Beta$, which are misaligned with the stream track especially in the leading arm (right side). Panel F (reproducing the bottom panel of Figure~\ref{fig:misalignment}) illustrates this misalignment in a different way: solid red line shows the ratio of proper motion components, while dashed black line is the slope of the mean $\Beta$ as a function of $\Lambda$. Systematic uncertainties on the reflex-corrected proper motions are illustrated by a red shaded region. Milky Way disc plane is shown by green line in panel G, and its centre -- by a green dot; the solar position is marked by a red star, and the direction towards the remnant -- by a red arrow.
} \label{fig:stream_obs}
\end{figure*}

\begin{figure*}
\includegraphics{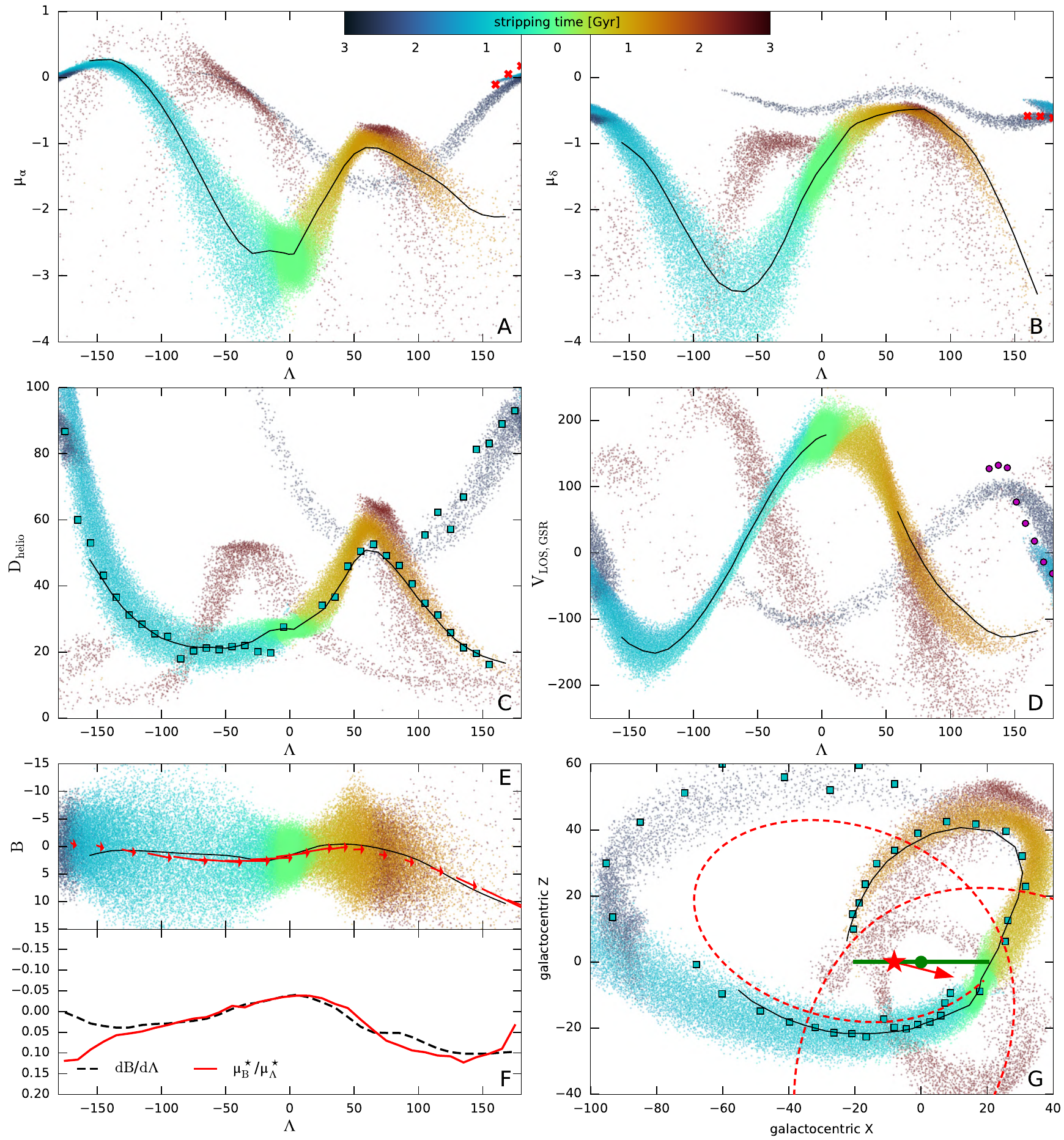}
\caption{Observable properties of the Sgr stream in a fiducial model without the LMC, but with a twisted non-triaxial Milky Way halo. Stellar particles are coloured according to their stripping time, defined as the most recent time when a particle left the sphere of radius 5 kpc around the progenitor. The remnant is coloured in green, trailing arm -- in blue/cyan, leading arm -- in orange/red. Panels are the same as in Figure~\ref{fig:stream_obs}, and the observed trends are shown by solid black lines in both figures. Panel F shows that in a static potential, even with a rather complicated shape, there is virtually no misalignment between the slope of the simulated stream track (dashed black) and the reflex-compensated proper motions (solid black). Dashed red line in panel G shows the past orbit of Sgr.
} \label{fig:stream_model_nolmc}
\vspace*{1cm}
\end{figure*}

\begin{figure*}
\includegraphics{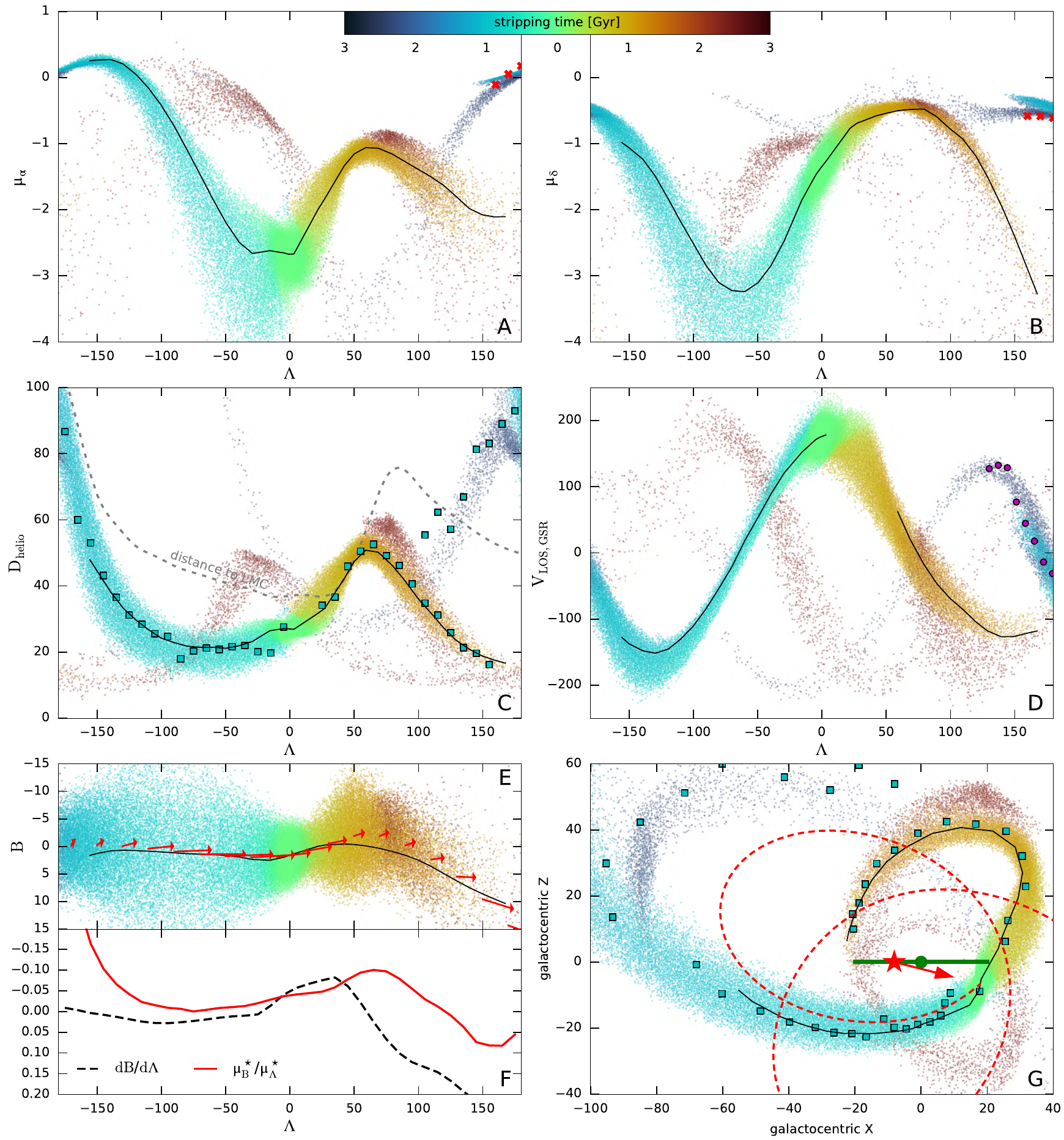}
\caption{Observable properties of the Sgr stream in our fiducial model with the LMC mass of $1.5\times10^{11}\,M_\odot$. Panels are the same as in the previous figure. Note the much better match for the distances in the leading arm apocentre in panel C, line-of-sight velocities in the Northern hemisphere ($\Lambda_\mathrm{Sgr}>0$) in Panel D, and a significant misalignment between the slope of the simulated stream track (dashed black) and the reflex-compensated proper motions (solid black) in Panel F, which qualitatively resembles the observations.
Panel C additionally shows the distance of the closest approach to the LMC (gray dashed line). The Sgr remnant and adjacent sections of both arms have the closest encounter $\sim100$~Myr ago at a distance $\sim40$~kpc, and particles around apocentre of the leading arm and further away from the remnant -- at significantly larger distances $\sim60-70$~kpc, yet they are deflected more strongly. See \url{https://youtu.be/dy_9GGMIXkU} for a movie of Sgr's disruption in this fiducial setup.
} \label{fig:stream_model}
\vspace*{1cm}
\end{figure*}

\section{Fitting the Sgr stream}  \label{sec:fit_results}

To examine the performance of our models, we will compare projections of the simulated tidal debris with the measurements in hand. A summary of the observed properties of the Sgr stream is presented in Figure~\ref{fig:stream_obs}. The top two panels, A and B, give the components of the proper motion $\mu_{\alpha}$ and $\mu_{\beta}$, correspondingly, as functions of the stream longitude $\Lambda$. Here and in the panels below, solid black line represents the observed stream track estimated by fitting cubic splines to the measured values of individual stars. The stream proper motion sample assembled in this work is complemented with the measurements of the trailing arm wrap presented in \citet{Ibata2020}. Next, panel C compares the run of heliocentric stream distances as dictated by the \Gaia RR Lyrae (see above) with the earlier measurements of \citet{Hernitschek2017}. In the same row, panel D shows the heliocentric velocities we procured by cross-matching our RGB member catalogue with public spectroscopic datasets as well as the measurements of the velocities in the trailing arm from \citet{Belokurov2014}. Panel G converts the stream positions and distances into Galactocentric $X$ and $Z$. Finally, panel E deals with the stream track on the sky, while F compares the stream direction (dashed black) with the proper motion direction (solid red), replotting the lower panel of Figure~\ref{fig:misalignment}. Below, we will present our stream models using the same arrangement of panels.

\subsection{Overall strategy}  \label{sec:workflow}

Our fitting approach has several steps. We first explore the parameter space of the Milky Way halo shape and orientation and the LMC mass with the MCMC approach, using the restricted $N$-body simulation (Section~\ref{sec:fast_stream_generation}) to generate the stream for each choice of parameters (stage A). This allows us to identify the region of the parameter space with reasonably looking models (if there are any). We then may examine selected models more closely by running a full $N$-body simulation of a disrupting Sgr galaxy in the given potential. In doing so, we still have two choices: either to keep the Milky Way and LMC potentials fixed to their initial profiles and retain the approximate trajectory of the LMC computed using Equation~\ref{eq:lmc_orbit}, as in the restricted $N$-body simulations, or to run a full $N$-body simulation of the Milky Way--LMC interaction and represent their evolving potentials by multipole expansions (stage B). Of course, this stage B is only relevant for models with the LMC, otherwise we just happily use the static Milky Way potential all the way through. In either case, several iterations of live Sgr simulations (stage C) are needed to arrive at the correct present-day position and velocity and a reasonable mass of the remnant (judged by comparing the simulated kinematic maps with observations, as in \citetalias{Vasiliev2020}).

Stage A (MCMC analysis of a large number of models with restricted $N$-body simulations) is, in some sense, preliminary, and we ultimately assess the quality of the fit using the full $N$-body models (stage C). As discussed below, there are some subtle but noticeable differences in the stream properties generated by the two approaches, especially if we use evolving Milky Way and LMC potentials (i.e., having an intermediate stage B) as opposed to static ones. Nevertheless, the conclusions about the potential parameters obtained from the restricted $N$-body simulations are unlikely to be seriously affected by the more approximate nature of this method.

\subsection{Static Milky Way}  \label{sec:no_LMC}

We first explore the possibility of fitting the Sgr stream by a static Milky Way potential, ignoring the presence of the LMC, as done in almost all previous studies. We run a series of MCMC searches for Milky Way halo models of increasing complexity, starting from a spherical one (which has 4 free parameters -- scale radius $r_\mathrm{h}$ and three dimensionless coefficients $\alpha,\beta,\gamma$ controlling the radial density profile), then allowing it to have a variable flattening in $z$ (adding the inner and outer axis ratios $q_\mathrm{in}$, $q_\mathrm{out}$ and the transition radius $r_q$), and finally allowing the outer halo to have three different principal axes with an arbitrary orientation relative to the disc, while keeping the inner part axisymmetric and aligned with the disc plane (this adds another axis ratio $p_\mathrm{out}$ and three angles specifying the orientation). Only in the latter case were we able to find configurations that reproduce most properties of the stream. 

The best-fit halo shape and orientation is indeed rather peculiar: the inner part of the halo is moderately oblate, with axis ratio $z:R \simeq 0.5-0.6$, while the outer part is strongly prolate and misaligned with the principal axes of the disc.  It has long been recognised that the halo needs to be non-spherical in order to produce the observed morphology of the leading arm: in a spherical halo, it bends much more strongly, and the stream crosses the Galactic plane around the Solar radius, not at $\sim 10-15$~kpc from the Sun as warranted by observations \citep[see e.g.][]{Belokurov2006,Yanny2009}. Allowing the halo to be prolate (extending more along the $z$ axis) in the outer part ``unbends'' the leading arm and moves its disc crossing point further out, while the additional flexibility allowed by misaligned principal axes improves the fit of the stream track. At the same time, the oblate inner part aligned with the disc shifts the distant portion of the trailing arm up (closer to the Galactic plane), improving the fit for both the distance and velocity at $\Lambda<-100^\circ$.

Figure~\ref{fig:stream_model_nolmc} shows a fiducial model from this series (static Milky Way in the absence of the LMC) projected into the observable space (proper motions in panels A and B, distance in panel C, line-of-sight velocity in panel D, and stream track in panel E). It can be directly compared with observations shown in Figure~\ref{fig:stream_obs}; the mean observational trends are overplotted by black lines in both figures. The agreement between the model and the data is very good in the trailing arm, but the apocentre of the leading arm in the model is too far (at $55-60$~kpc) compared to the observed $\sim 50$~kpc, and there is some mismatch in the line-of-sight velocities around the apocentre of the trailing arm. Crucially, there is no misalignment of the reflex-corrected proper motions from the stream track (panel F), as indeed expected in a static potential. This is in marked contrast with the observations, which show a significant misalignment in the leading arm. We thus conclude that such a model is unable to match all observed features of the stream.

\subsection{Moving Milky Way + LMC}  \label{sec:with_LMC}

\begin{figure*}
\includegraphics{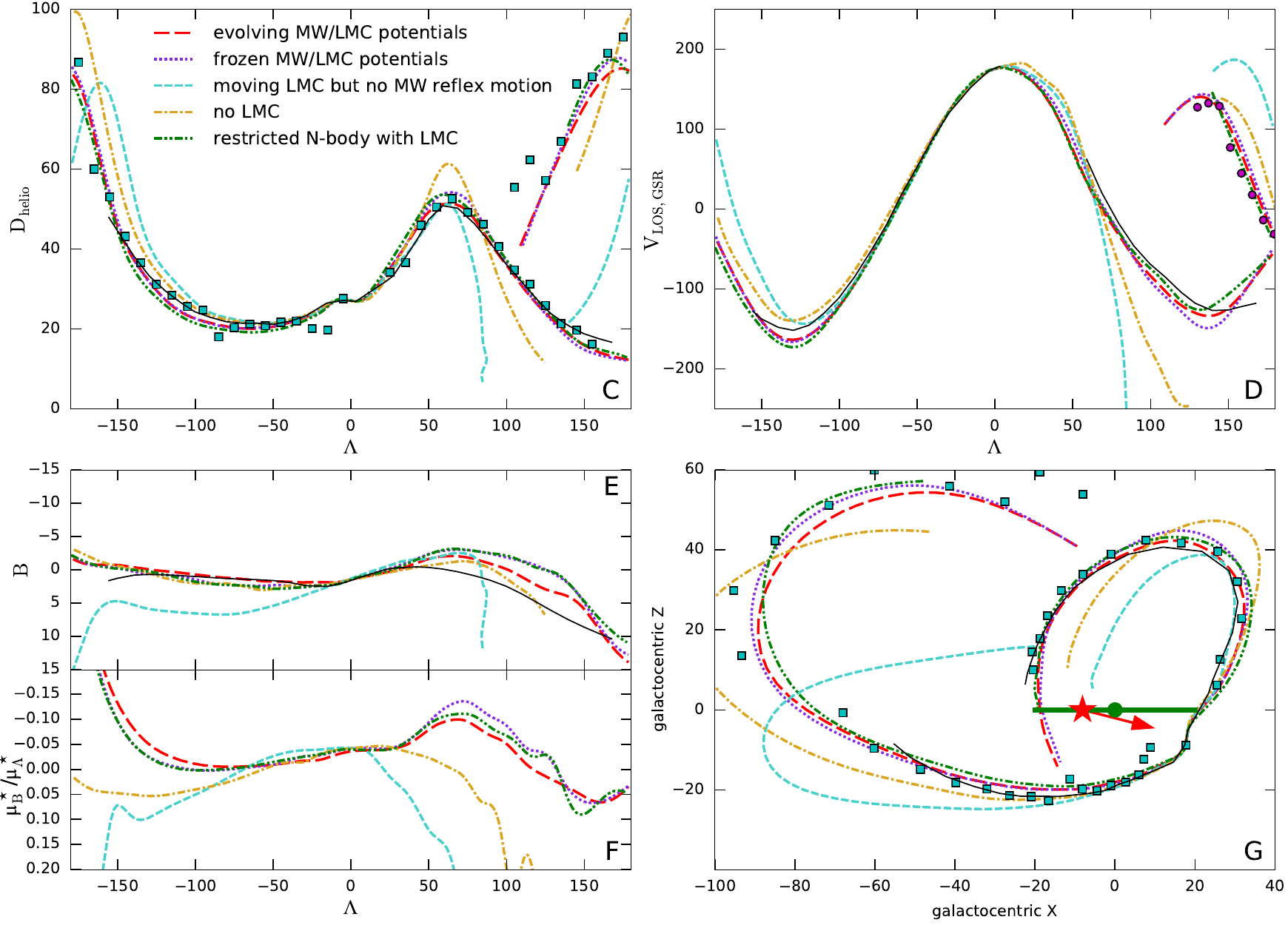}
\caption{Comparison of stream models in the same Milky Way potential, but with different dynamical mechanisms selectively turned on or off. Shown are the same quantities as in the previous figures (distance, line-of-sight velocity, stream track, and Galactic side-on view), except proper motions (these do not show any features not evident from other panels). For each model, only the mean trends are shown with coloured lines, and the data trends are plotted as a black solid line and boxes or circles. Red long-dashed curves show the baseline model (same as in Figure~\ref{fig:stream_model}) with $M_\mathrm{LMC}=1.5\times10^{11}\,M_\odot$, and both Milky Way and LMC potentials represented as evolving multipole expansions. Blue dotted curves show the model in which the Milky Way and LMC potentials are frozen to their initial (unperturbed) profiles; it deviates slightly but noticeably in the apocentre distances and in the stream track. Cyan short-dashed curves show an unphysical model in which the moving LMC potential is taken into account, but the corresponding acceleration of the Milky Way-centered reference frame is neglected; this clearly produces a completely wrong stream trajectory. Orange dot-dashed curves show the stream in the same Milky Way potential without any influence from the LMC; of course, it deviates from the actual data much more strongly than the best-fit model \textit{in a different potential} without the LMC (Figure~\ref{fig:stream_model_nolmc}), but still highlights the main features introduced by the LMC (un-bending of the leading arm and an upward shift of the trailing arm). Finally, green dash-double-dotted curves show the restricted $N$-body simulation (stage A), which follows the live Sgr in a frozen Milky Way+LMC potential in its overestimate of the apocentre distances. All other curves depict live $N$-body models (stage C). A more detailed comparison between the second and the last curves is presented in Figure~\ref{fig:stream_compare_restricted}.
} \label{fig:stream_model_comparison}
\end{figure*}

We now consider the situation when a massive LMC flies by the Milky Way, and the latter acquires reflex-motion velocity in an inertial reference frame. As discussed in Section~\ref{sec:Nbody_simulations}, we find it is more convenient to work in the Milky Way-centered non-inertial frame and add a spatially uniform time-dependent acceleration field to the equations of motion for the Sgr system, as well as the gravitational field from a moving LMC. As before, we first conduct a series of MCMC searches with variable halo properties, also adding the LMC mass as another free parameter. We find that the observational properties of the stream are best reproduced by models with $M_\mathrm{LMC}$ in the range $(1-1.7)\times10^{11}\,M_\odot$, and pick $M_\mathrm{LMC}=1.5\times10^{11}\,M_\odot$ as our fiducial value for further exploration.

The effect from adding the LMC is, to some extent, similar to the effect from making the Milky Way halo prolate -- it ``unbends'' the leading arm and brings its apocentre closer. At the same time, it shifts the nearest part of the trailing arm to slightly smaller distances and larger (more negative) line-of-sight velocities, spoiling the perfect agreement seen in the models without the LMC. We find that the latter effect may be somewhat compensated by allowing the halo to be oblate in the inner part and transition to a moderately prolate shape in the outer part, while keeping it axisymmetric and aligned with the principal axes of the disc. Slightly better fits were obtained by allowing the halo to be triaxial in the outer part, but with two of its axes ($X$, $Y$) still lying in the disc plane. Figure~\ref{fig:stream_model} shows a fiducial model from this series, projected into the observable space in the same way as before. While this particular model is still not a perfect match to the data, the key deficiencies of the LMC-less model have mostly been rectified. The distances and velocities in the leading arm are much closer to the data, as well as the velocities near the apocentre of the trailing arm. Most notably, the stream track on the sky is now misaligned with the direction of reflex-corrected proper motions (panels E and F) in the leading arm, even though the stream track is slightly offset from the observations. The magnitude of the misalignment is roughly proportional to the LMC mass, but also depends somewhat on the halo geometry.

The agreement between the restricted and full $N$-body simulations (stages A and C) is worse in the case of a large and extended LMC. The former method still produces a qualitatively similar stream, but the apocentric distances are overestimated, making it difficult to reliably explore the parameter space. We find that this disagreement is driven, at least partially, by the distortions in both the Milky Way halo and the LMC itself in the fully self-consistent simulation at stage B. If we neglect the perturbations of the Milky Way and LMC potentials and keep them frozen at the initial profiles (of course, with the LMC still moving along its actual trajectory), jumping from stage A directly to C, the ``live'' Sgr simulation resembles more closely the restricted $N$-body simulation. Figure~\ref{fig:stream_model_comparison} illustrates this slight difference between evolving and frozen potentials (red and blue curves, correspondingly); the stream generated by a restricted $N$-body simulation is shown by green curves and is similar to the live Sgr model in frozen Milky Way+LMC potentials. We also see that if we do not take into account the Milky Way reflex motion (cyan curve), this drastically changes the shape of the Sgr stream, highlighting the fundamental importance of this factor (as stressed by \citealt{Gomez2015}, \citealt{Petersen2020}). Of course, it is also evident from the fact that the effect of the LMC is seen most dramatically in the leading arm (northern Galactic hemisphere), while the LMC itself flew by in the southern hemisphere. Finally, a simulation in the same Milky Way potential but without any LMC influence (orange curve in Figure~\ref{fig:stream_model_comparison}) illustrates the main difference in the stream morphology -- larger apocentre distances and higher orbit eccentricity in the LMC-less scenario.

As explained in Section~\ref{sec:Nbody_simulations}, we approximately take into account dynamical friction acting on the Sgr galaxy by adding an extra drag term in the equations of motion of the live $N$-body system evolved in a pre-recorded background potential. For the initial Sgr masses considered in our models (up to $4\times10^9\,M_\odot$), neglecting dynamical friction would cause a decrease of the apocentre distance for the trailing arm by $\sim 3$~kpc, and for the leading arm by $\sim 1$~kpc: a relatively minor (but still noticeable) change. Other observable properties of the stream are largely insensitive to dynamical friction.

\subsection{Properties of the remnant and the stream}  \label{sec:remnant}

\begin{figure*}
\includegraphics{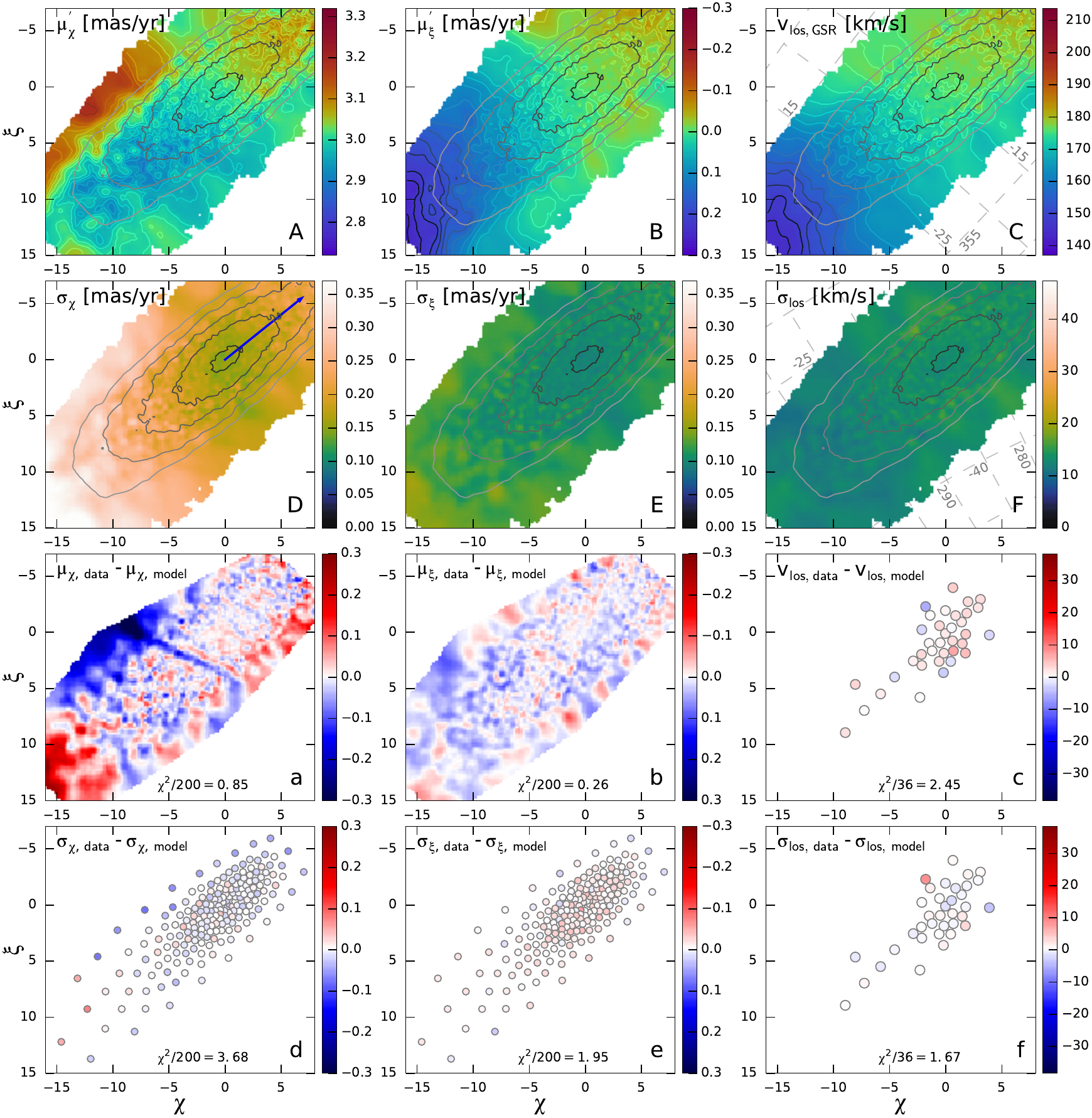}
\caption{Kinematic maps of the Sgr remnant in our fiducial model. Coordinates are aligned with the apparent (non-reflex-corrected) motion of the remnant on the sky (the motion is in the direction of increasing $\chi$) for reasons explained in Section~5.1 of \citetalias{Vasiliev2020}. The actual velocity vector of the Sgr galaxy is shown by a blue arrow in panel D and points roughly towards the Galactic plane, which lies slightly beyond the top right corner. A grid of Galactic coordinates is shown in panel C, and equatorial coordinates -- in panel F. \protect\\
Panels A and B show the two components of perspective-corrected mean proper motion, panels D and E -- their dispersions; Galactic standard-of-rest line-of-sight velocity is displayed in panel C, and its dispersion -- in panel F. These plots can be directly compared to the actual observations, shown in Figure~5 of \citetalias{Vasiliev2020}; the differences between the model and the data are plotted in the corresponding panels of two bottom rows.
} \label{fig:remnant_model}
\end{figure*}

In all these fitting exercises, we tuned the initial radius of the Sgr progenitor so that it loses just the right amount of mass by the present moment. We find that the present-day total mass within 5~kpc from the Sgr centre needs to be in the range $(3-4)\times10^8\,M_\odot$ in order to satisfy the velocity dispersion constraints. Figure~\ref{fig:remnant_model} illustrates the properties of the remnant, projected into the same observable space as Figure~5 in \citetalias{Vasiliev2020}, and can be directly compared with that figure, which shows the observed mean proper motion and line-of-sight velocity and their dispersions across the face of the galaxy. The agreement is satisfactory, even though there are some deficiencies.

\begin{figure}
\includegraphics{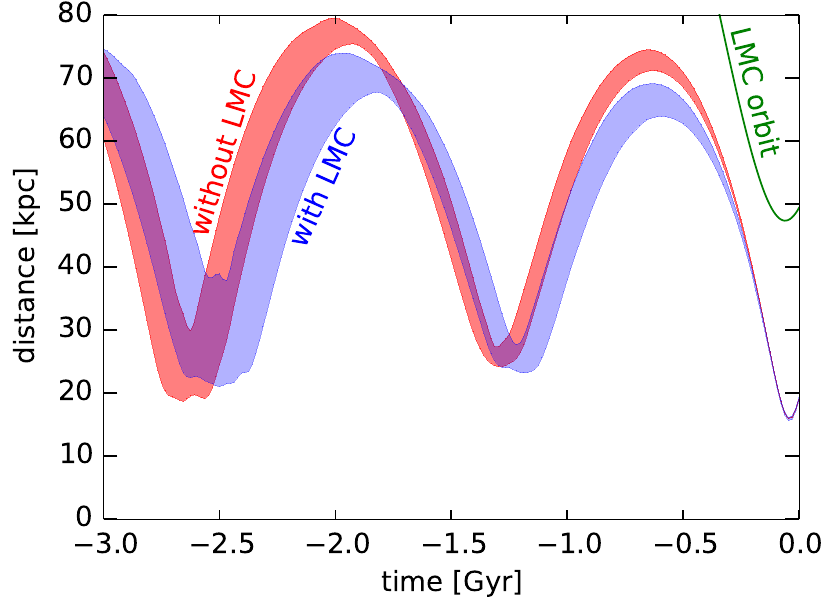}
\includegraphics{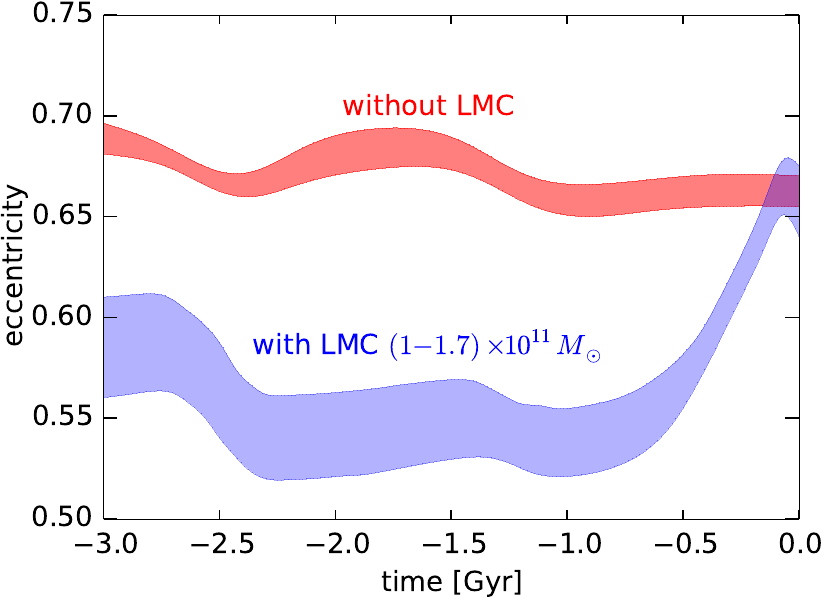}
\caption{
Evolution of galactocentric distance (upper panel) and eccentricity (lower panel) of Sgr orbits from the ensemble of MCMC simulations with (blue) and without (red) LMC. The orbits coincide over the last $\sim 200$~Myr, being constrained by the present-day position and velocity of the Sgr remnant. However, in models with the LMC the orbit has a slightly shorter period and lower eccentricity for most of its evolution, and increases eccentricity just recently, deflected by the LMC. The apocentre decays due to dynamical friction, while the pericentre changes non-monotonically due to non-sphericity of the halo potential. Green curve in the top panel shows the galactocentric distance of the LMC, which happens to pass through its pericentre around the same time as Sgr, $\sim50$~Myr ago.
} \label{fig:trajs}
\end{figure}

The top left panel showing the mean proper motion along the apparent (non-reflex-corrected) direction of motion demonstrates a prominent blue ``dip'' of lower values in the lower right corner, which then transitions to higher (red) values at larger distances from the Sgr centre. As argued in the above paper, this feature corresponds to the trailing part of the remnant being located at larger heliocentric distances, and the transition from the remnant to the stream reverses this distance gradient and increases the proper motion values. The distance from the Sgr centre corresponding to this transition is somewhat larger in our models than the data warrants. This is in contrast to the findings of \citetalias{Vasiliev2020}, which had the opposite difficulty of extending the transition radius as far as the data indicates. The slightly increased (by less than 2\%) heliocentric distance to the remnant in the present study apparently has a disproportionally large influence on the spatial extent of the remnant.

Another aspect of minor disagreement between the model and the observation is the orientation of the remnant with respect to the stream. Dashed black curve in panel F of Figure~\ref{fig:stream_obs} shows the inclination angle (the gradient of mean $\Beta$ as a function of $\Lambda$), which demonstrates a prominent kink around the location of the remnant ($-30^\circ < \Lambda < 10^\circ$). This gradient is much smoother in our models (e.g., panel F of Figure~\ref{fig:stream_model}). The apparent difference between the two figures is greatly exaggerated by the aspect ratio of the plot, and the actual mismatch between the major axis orientations is only $2^\circ$. We find that it is pretty similar between all models with a spherical progenitor, but can be affected if we assume an initially rotating Sgr model (even with a rather slow one with the peak rotation velocity smaller than the central velocity dispersion). The adjacent parts of the Sgr stream are also somewhat affected by the rotation in the progenitor, and this mechanism has been invoked to explain the bifurcation in the stream \citep{Penarrubia2010}, although that particular model does not match current observational constraints. A full exploration of the parameter space of rotating models is beyond the scope of this paper, and we focus on the simpler case of a spherical non-rotating progenitor in the rest of the analysis.

By construction, the Sgr galaxy in our models experiences only 3 pericentre passages over the course of the simulation, with the most recent one just $\sim40$~Myr ago. Over this time, it has lost up to $90\%$ of the initial dark matter mass, and approximately half of the stellar mass (the latter fraction matches the observational constraints from \citealt{NiedersteOstholt2010}). The tidal stripping of stars only began in earnest during the previous pericentre passage $\sim1.2$~Gyr ago, and the entire trailing arm up to its apocentre is formed from this material stripped at that time (shown by cyan in Figure~\ref{fig:stream_model}). The apocentre of the trailing arm and its further extension in the Galactic northern hemisphere are stripped $\sim2.5$~Gyr ago (darker blue). By contrast, the leading arm is populated by the material stripped at the previous passage (orange) over its entire arc in the northern hemisphere; the material stripped earlier (darker red) is spatially and kinematically offset and forms a more distant plume around the apocentre, extending back towards the Galactic disc and possibly further into the southern hemisphere. To summarize, almost the entire leading arm and the trailing arm up to its apocentre were stripped approximately at the same time (during the previous pericentre passage). By contrast, in the \citetalias{Law2010a} model the progenitor had a lower mass and the tidal stripping proceeded much more gradually over several pericentre passages, producing a stronger gradient of stripping time along the stream.

We also highlight the dramatic contrast between the Sgr orbit (dashed red curve in panel G) and the stream track, with $\sim 50\%$ difference in the apocentre distance of the trailing arm. This deviation of the stream from the orbit is stronger in our models than in the \citetalias{Law2010a} model, and is again a consequence of a much larger progenitor mass. Moreover, the stream has been deflected by the LMC in the last 0.5~Gyr, whereas the orbit recorded at earlier times is, of course, unaffected. However, the LMC increases the offset of the stream from the orbit only slightly, as seen by comparing Figures~\ref{fig:stream_model_nolmc} and \ref{fig:stream_model}. The main effect of the LMC on the orbit is the decrease of the initial eccentricity (Figure~\ref{fig:trajs}, lower panel) and a slight increase of the orbital period (upper panel). In other words, the Sgr orbit in the presence of the LMC must be less eccentric initially in order to match the present-day position and velocity of the Sgr remnant, which have been dramatically affected in a short time interval just recently. If continued along its unperturbed orbit, Sgr would have missed its present-day position by $\sim 8$~kpc and velocity by $\sim 30$~\kms.

Note that the orbital timescales in our simulations are considerably longer, and apocentre distances -- larger than in some recent studies exploring the effect of the Sgr on the Milky Way disc (e.g., \citealt{Purcell2011,Laporte2018,BlandHawthorn2020}), calling for a reanalysis of its role in seeding the phase-space spiral in the Milky Way discovered by \citet{Antoja2018}. The previous disc crossing in our MCMC series occurs $\sim 1.2-1.3$~Gyr ago at a distance $\sim 24-25$~kpc, and the Sgr mass is $\lesssim 2\times10^9\,M_\odot$ at this moment. Interestingly, these longer timescales are in agreement with the recently measured star formation bursts in the Milky Way disk, which have been associated with Sgr in \citet{RuizLara2020}. These constraints on Sgr's orbital history will also help determine whether Sgr has perturbed other tidal streams in the Milky Way \citep[e.g.][]{deBoer2020,Li2020}. While our simulations do not have any hydrodynamic component, the presence of gas in the Sgr progenitor might have affected its orbital history, and the pericentre passages are expected to leave imprint in the star formation history of the dwarf \citep[e.g.][]{TepperGarcia2018}.

The apocentre of the trailing arm is $\sim10\%$ closer (but within the error-bars) in our models than the $\sim100$~kpc measured from RR Lyrae in the PanSTARRs catalogue \citep{Hernitschek2017}, although it roughly matches the smaller distances inferred by \citet{Ramos2020} for the \Gaia RR Lyrae. However, the stream is relatively thick and consists of two superimposed stripping episodes in this region, so this mismatch might be ameliorated by modifying the progenitor properties responsible for the mass loss history.  Our models do not reproduce the bifurcations seen in the Sgr stream (e.g., \citealt{Belokurov2006}, \citealt{Koposov2012}), nor are they designed to. These morphological features, age and metallicity gradients along the stream and various other observed properties are outside the scope of this study.

The larger apocentre distance in our models compared to the \citetalias{Law2010a} model unambiguously associates the enigmatic globular cluster NGC~2419 with the stream, matching not only its distance, but also line-of-sight velocity and proper motion. This association has been long conjectured based on the observed stream properties (e.g., \citealt{Belokurov2014}, \citealt{Sohn2018}), but is now shown to be supported by models. We also confirm the association of two clusters in the trailing arm (Pal~12 and Whiting~1), as well as four clusters in the remnant (M~54, Arp~2, Terzan~7 and Terzan~8); however, no other clusters considered by \citet{Law2010b} and \citet{Bellazzini2020} appear to be anywhere close to the stream, if we consider all 6 phase-space dimensions. Note, however, that our models are only evolved for the last 3~Gyr, and make no predictions about the structure of more ancient stream wraps.

\section{Discussion}  \label{sec:discussion}

\begin{figure*}
\includegraphics{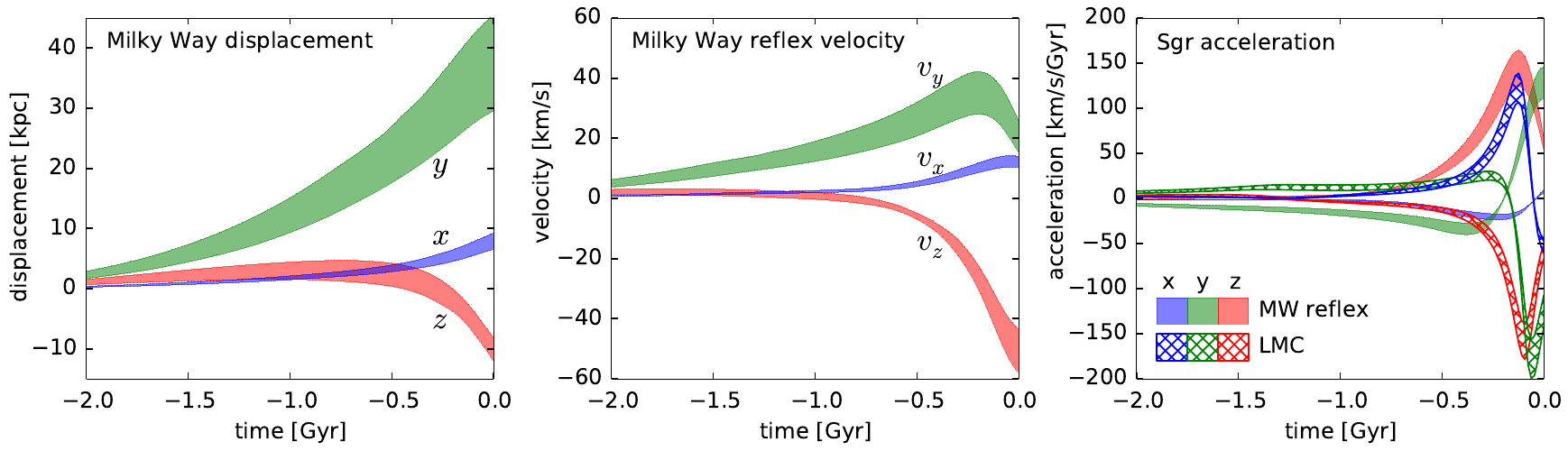}
\caption{Illustration of the importance of Milky Way reflex motion.
\textit{Left} and \textit{centre panels} show changes in the Milky Way position and velocity in response to the LMC flyby as a function of time, for the MCMC series with a variable LMC mass. The offsets are measured in the inertial reference frame, in which the Milky Way position and velocity are set to zero at the beginning of the simulation ($-3$~Gyr). Note that the velocity change is the same in all inertial frames, but the position change is not. Simulation of the Sgr stream are performed in a non-inertial frame centered at the Milky Way at all times, and the corresponding time-dependent acceleration (time derivative of the Milky Way reflex velocity with inverse sign) is added to the equations of motion. \textit{Right panel} shows the contribution of this reflex-induced acceleration (shaded curves) and the acceleration from the LMC itself (hatched curves) to the total acceleration of the Sgr remnant as a function of time. The two time-dependent acceleration sources are of the same order and mostly (but not entirely) cancel each other. The acceleration from the Milky Way potential itself is $\sim 10\times$ larger than both these additional components.
}  \label{fig:reflex}
\end{figure*}

\subsection{The dance of three galaxies}

The most unambiguous signature of the strong gravitational potential perturbation (likely) induced by the LMC is the apparent misalignment between the stream track and the direction of its reflex-corrected proper motion (see e.g. Figure~\ref{fig:misalignment}). Ours is not the first observation of such a deviation (see \citealt{Koposov2019,Erkal2019} for the measurement and the modelling of the Orphan Stream deflection by the LMC and \citealt{Shipp2019,Li2020} for proper motion offsets observed in streams in the Southern hemisphere). In the Sgr stream, as we argue above, in addition to the velocity misalignment, there are other rather subtle tell-tale signs, for example the ``unbending'' of the leading arm. It is not immediately clear how the interplay of the additional forces due to the LMC's pull and the accelerated motion of the Milky Way affected the stream. To address this question, in this Section we attempt to further elucidate the mechanics of the interaction between the Sgr stream with the Milky Way in the presence of the LMC.

Let us re-iterate that it is the leading arm of the stream that appears to be perturbed the most. The velocity misalignment reveals that the leading arm has been tugged in the direction perpendicular to the Sgr motion, i.e. along $\Beta$ which is roughly aligned with the Galactocentric $y$-axis and with the LMC's orbital plane. The LMC however does not move only along $y$: as it approaches the Galaxy, the Cloud has a substantial component of its motion along $z$, which may explain the reshaping of the loop of the leading tail apparent on comparison between Figure~\ref{fig:stream_model_nolmc} and \ref{fig:stream_model}. Note however, that unlike the case of the Orphan, it is not the direct deflection of the orbits of the Sgr stars by the LMC's gravity that distorts the stream. Rewinding the simulations back in time, it is clear that in the part of the stream near the progenitor (i.e. with $|\Lambda| <120^{\circ})$, the closest approach to the Cloud is achieved by the stars in the trailing arm as illustrated with the dashed grey line in the panel C of Figure~\ref{fig:stream_model} (excluding the unobserved portion of the leading arm with $0<\Lambda<40$). In fact, as Figure~\ref{fig:stream_model_comparison} demonstrates, the action of the LMC alone (i.e. in the unphysical case of the static Milky Way) is to pull the stream in the opposite direction (cyan short-dashed line). The Cloud drags the stream stars toward itself, to negative $\Beta$ and negative $z$, while the observations indicate a clear additional motion to positive $\Beta$ and slight ``unbending'' towards positive $z$.

We therefore conclude that the action of the accelerated Milky Way is needed to balance the pull of the LMC. The Milky Way reflex motion is further illustrated in Figure~\ref{fig:reflex}, which shows the changes in the Galaxy's position and velocity over the course of the simulation (a few tens of kpc and \kms, comparable to figure~3 in \citealt{Gomez2015}), as well as the corresponding accelerations in the motion of the Sgr dwarf. In line with the LMC's orbital geometry, the largest kicks the Galaxy experiences are along the $y$ and $z$ directions. More precisely, by the present day, the Milky Way has shifted the most along $y$ by some 30 kpc, but has acquired the largest impulse along $z$ of order of $-50$ kms$^{-1}$. The $x$-axis is almost perpendicular to the Magellanic orbital plane, thus displacement along this direction is minimal (left and middle panels of the Figure). As the right panel reveals, the non-inertial accelerations due to the moving Milky Way are of the same order as the acceleration from the LMC itself, but opposite in sign, so largely cancel out (which explains the dramatic difference in the stream behaviour if we omit the reflex motion-induced acceleration). Note that the synchronization of the two forces is not complete and depends on the stream latitude; as a result, the acceleration balance is imperfect. While along a large portion of the trailing arm, the Milky Way's reflex appears to have compensated the LMC's pull, in the leading arm, which is further from the Cloud, the Milky Way's non-inertial acceleration (which is the same throughout the Galaxy) overpowers the action of the Cloud and starts to bend the tail away from the LMC.

\begin{figure*}
\includegraphics{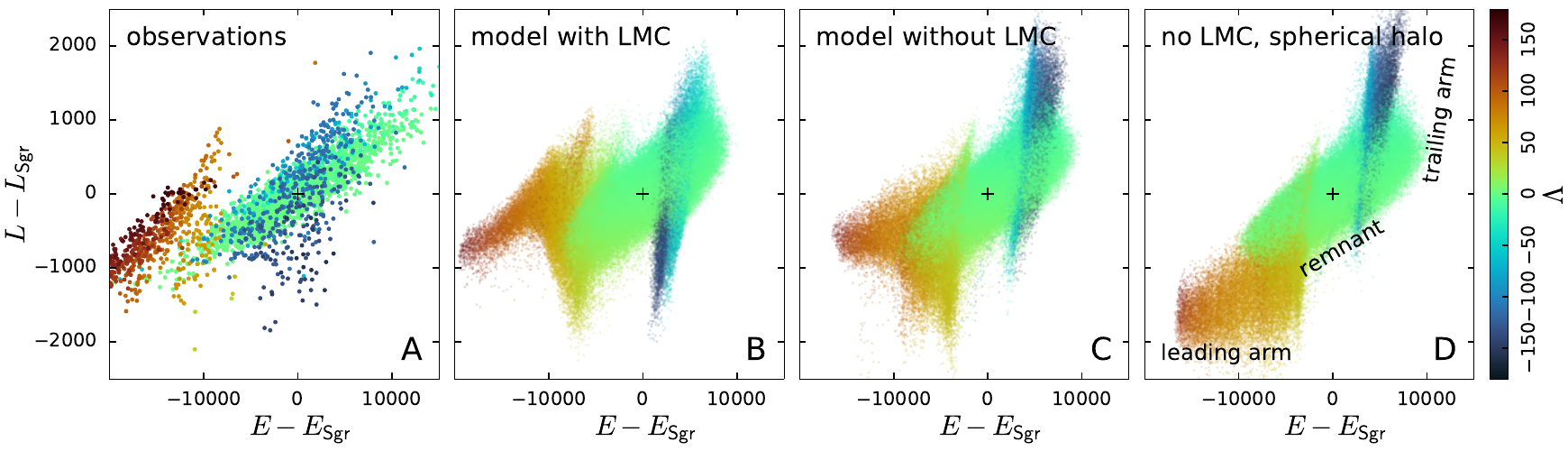}
\caption{Total angular momentum $L$ vs.\ energy $E$ in the observations (panel A), model with the LMC (panel B), and two models without LMC: with a twisted non-spherical halo (panel C) and with a spherical halo (panel D), coloured by the stream longitude $\Lambda$. Energy is measured in units of (\kms)${}^2$, and angular momentum -- in \kms\,kpc$^{-1}$; both are shifted to place the remnant at zero (marked by black crosses). In panel A, we only plot the stars with 6d phase-space coordinates, and for each star, average the proper motions over 20 nearest stars on the sky (including those with only 5d coordinates) to reduce the noise. For a stream in a static spherical potential (panel D), both the energy and the angular momentum the leading (trailing) arm are lower (higher) than the remnant. In a static non-spherical potential (panel C), the angular momentum is not conserved and its variation depends on the position; as a consequence, the leading arm shifts towards higher values of $L$. In the model with a strong perturbation from the LMC (panel B), the variation in $L$ is much stronger and shifts the leading arm further up, and the variation in energy reverses the order in the trailing arm (the more recently stripped particles closer to the progenitor, coloured in cyan, end up at higher energy than the more distant particles, coloured in dark blue). A similar trend of $E$ vs $L$ is also seen in the leading arm in observations (panel A), although the trailing arm has a large scatter in energy, making it difficult to discern any clear trend.
}  \label{fig:EL}
\end{figure*}

A concise representation of the entirety of a stream can be attained by plotting energies and angular momenta of the constituent debris. Such an AM--E plot (or an equivalent built with actions) assumes that both quantities are conserved, which is true for energy (or action) in a static potential, but holds only approximately for the angular momentum, which can evolve in aspherical potentials. AM--E distributions have been used widely in the theoretical stream literature to illustrate the mechanics of the tidal tail formation. Examples include (but are not limited to) Figure 3 of \citet{Helmi2000}, Figure 9 of \citet{Penarrubia2006}, Figure 10 of \citet{Eyre2011}, Figure 1 of \citet{Gibbons2014}. Only recently, thanks to the unprecedented quality of the \Gaia data, has it become possible to explore distributions of integrals of motion based on the actual observations of Galactic stellar streams \citep[e.g.][]{Li2019}. In the AM--E space, the tidal debris corresponding to the trailing and leading arms form two distinct clusters, offset in opposing directions from the cluster corresponding to the stars still bound to the progenitor. The progenitor cluster takes a shape of an ellipse (approximately), which rotates and changes its aspect ratio and size as a function of the orbital phase. For a non-circular satellite orbit, the AM--E distribution of the stars inside the progenitor is not constant (even in spherical potentials) because the gravitational potential experienced by the stars inside the satellite varies with time. The AM--E distributions of the stripped stars sample the AM--E distribution of the parent object, more precisely its state at the time of stripping. To escape into the trailing tail a star is required to have a higher energy and angular momentum, and conversely, the leading arm is composed from stars with lower energy and angular momentum compared to the parent.

Figure~\ref{fig:EL} compares the distributions of relative (with respect to the Sgr dwarf) angular momentum and energy of the Sgr stream and the remnant based on the observations presented here (panel A) with several stream models (panels B, C and D). In this Figure, the colour-coding is set according to the stream longitude $\Lambda$, correspondingly the remnant stars are green and stars in the trailing (leading) arm are blue (red). Note that for this Figure, only stars (and particles) with $|\Lambda|<150^{\circ}$ are shown to make a consistent comparison between the data and the models. Panel D presents the simplest case of a Sgr-like stream in a spherical halo. As expected, the leading tail particles have negative relative energy $\Delta E$ and negative relative AM $\Delta L$, i.e. are located underneath the cloud of progenitor stars. Correspondingly, the trailing debris are found at positive $\Delta E$ and $\Delta L$, even though the shape is different because the $\Lambda$ cut (see above) affects the two tails differently. Comparing panels A and D reveals a striking difference between the observed AM--E distributions and those predicted for a stream in a spherical halo. While some trailing stars reside above the progenitor, the bulk of the trailing tail distribution is clearly shifted down, so that it largely overlaps with the positions of the stars in the remnant, many more (than expected) trailing stars have negative relative angular momenta. The behavior of the leading AM--E distribution is even more bizarre: the entire cloud appears to be shifted up, to positive $\Delta L$ such that it occupies the same AM range as the progenitor stars. In view of the ideas of the tidal tail formation mentioned above, these offsets in the AM with respect to the progenitor imply significant evolution of the AM after stripping.

Naturally, precession (and nutation) in an aspherical potential may lead to shifts and shape changes of the debris AM distribution. This is indeed the case, as illustrated in the panel C of Figure~\ref{fig:EL}. Here, the evolution of the leading tail in a strongly aspherical gravitational potential (see next Section) leads to a significant AM shift.
While this complicated and twisted halo model goes some way to re-arrange the AM distribution and reach a better agreement with the data, in the absence of the LMC, the match is far from perfect. The leading tail is not moved sufficiently far up, towards positive $\Delta L$ while the trailing tail is not dragged far enough down to negative $\Delta L$. The best agreement can be seen in panel B, which presents the results for our fiducial model with a massive LMC and a moving Milky Way. As a result of the action of the in-falling Cloud, the AM distributions of leading and trailing arms are shifted to match those observed. It is also clear that, while leading tail is affected the most (shows the largest shift in $\Delta L$), the trailing debris has acquired a non-zero offset in angular momentum. From examination of Figures ~\ref{fig:stream_model} and ~\ref{fig:stream_model_comparison} it is evident that the part of the trailing arm that suffered the most the disturbance due to the Cloud is that beyond $\Lambda=-150^{\circ}$, i.e. lying at large distances from the Sgr dwarf and the Milky Way.

\begin{figure}
\includegraphics{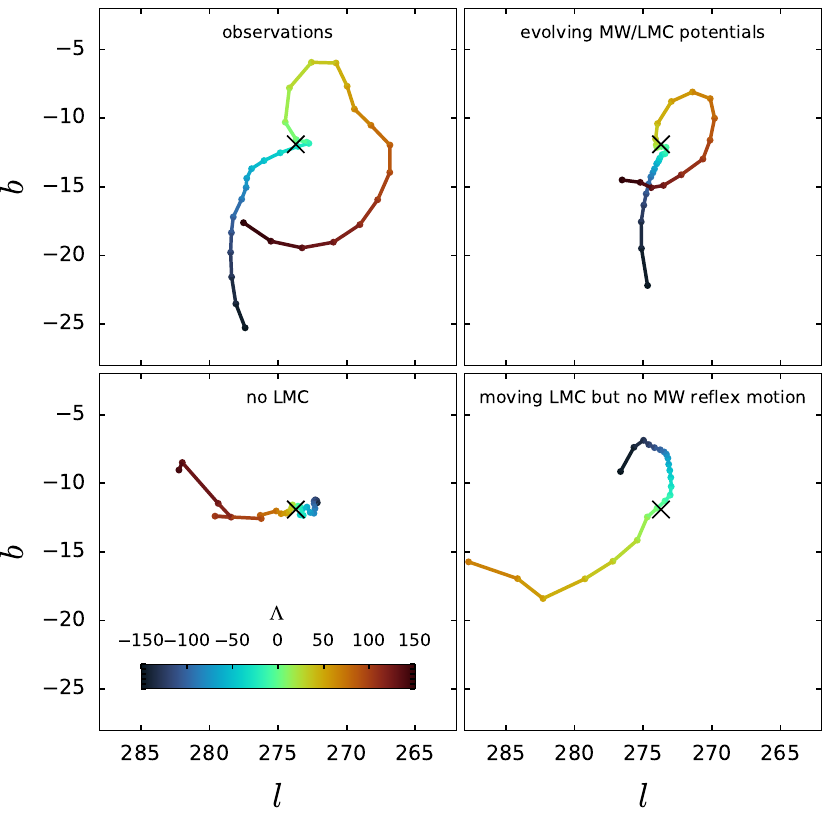}
\caption{Variation of the direction of angular momentum along the stream plotted in Galactic coordinates $l$, $b$. Top left panel shows actual observations, while other panels show models in the same Milky Way potential, but different mechanisms selectively turned on or off (same as in Figure~\ref{fig:stream_model_comparison}). Stream longitude $\Lambda$ is shown by colour, going from red in the leading arm through green in the remnant to blue in the trailing arm; dots are placed at every $10^\circ$. Black cross marks the angular momentum of the remnant. The model with the LMC reproduces the overall trends at least qualitatively, while other models have completely different behaviour. Note that this plot shows the kinematics, not the orbital plane of the stream -- the latter is quite similar between all models (panel E in Figure~\ref{fig:stream_model_comparison}).
}  \label{fig:angular_momentum_precession}
\end{figure}

Finally, an alternative view of the stream evolution in a non-trivial host potential can be provided by plotting the orientation of a plane passing through each segment of the debris, as has indeed been done previously for Sagittarius \citep[][]{Majewski2003, Johnston2005, Belokurov2014}. These studies found that the position of the debris plane pole evolves along the Sgr stream, likely due to differential precession (and nutation) in an aspherical potential \citep[see][]{Erkal2016}. Analogous but not completely equivalent to the debris plane evolution is the change of the angular momentum along the stream. In the simplest terms, this is because it would take the stars some considerable time to evolve away from their previous orbital configuration even if the torque was applied instantaneously. Such a prompt velocity kick would however be immediately registered in the value and orientation of the angular momentum. Before \Gaia, of course, it was simply unimaginable to be able to trace subtle angular momentum variations in the Milky Way outskirts. Thanks to the power of \Gaia astrometry and the large number of bona fide stream stars in our sample, we can now make such a plot. 

Figure~\ref{fig:angular_momentum_precession} shows the variation of the angular momentum direction along the stream, both for the actual observations (top left panel) and for several variants of models shown in Figure~\ref{fig:stream_model_comparison}. We stress again that this direction is, in general, different from the orbital pole of the stream -- the latter shows only the spatial orientation, while the former is a combination of the stream's spatial and kinematical properties. By construction, models that adequately fit the stream track on the sky and the heliocentric distance will also fit the orbital plane, and the three models shown on that figure do it reasonably well. However, only the model with the massive LMC and moving Milky Way is able to reproduce the observed trends at least qualitatively: the no-LMC model and the model without the reflex motion demonstrate the opposite trends to observations\footnote{\citet{Boubert2020} discuss another situation (the deflection of hypervelocity stars ejected from the Milky Way centre) and show that if the Milky Way reflex motion in response to the LMC flyby is neglected, it reverses the sign of the LMC's effect.}. Although the no-LMC model plotted on that figure (also shown by an orange curve in Figure~\ref{fig:stream_model_comparison}) uses the same potential as the best-fit model with the LMC, and hence gives a poor fit to the stream geometry, the angular-momentum track has the same qualitative behaviour in all model in a static potential, including our fiducial model shown in Figure~\ref{fig:stream_model_nolmc}.

\subsection{Inference on the Milky Way potential}  \label{sec:potential}

\begin{figure*}
\includegraphics{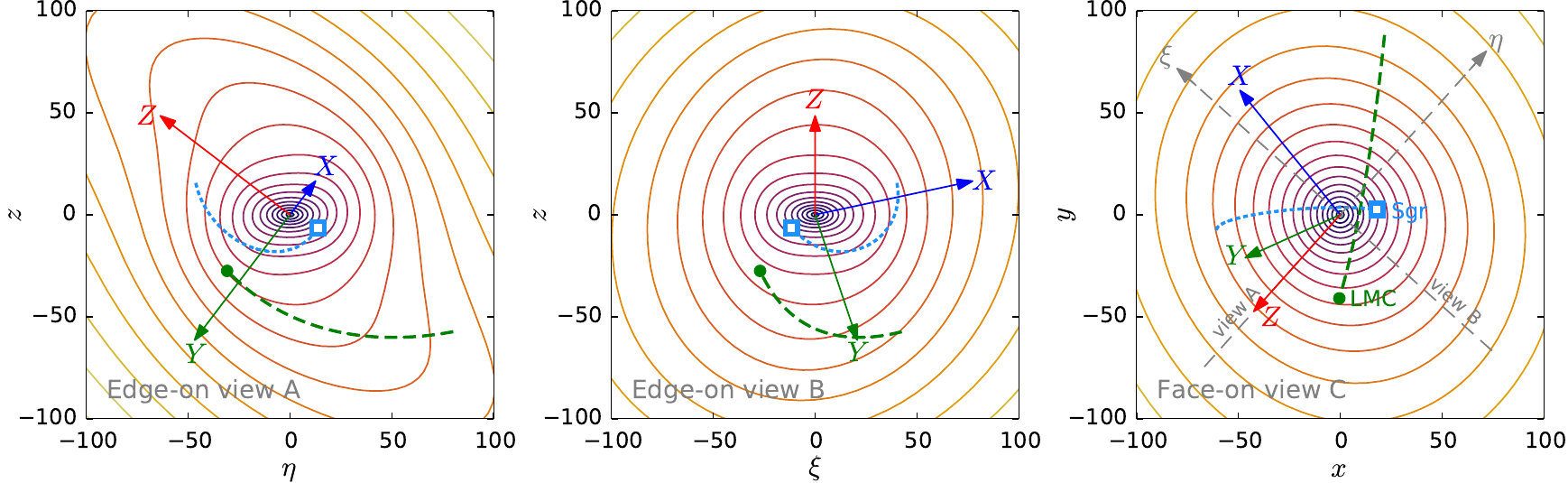}
\includegraphics{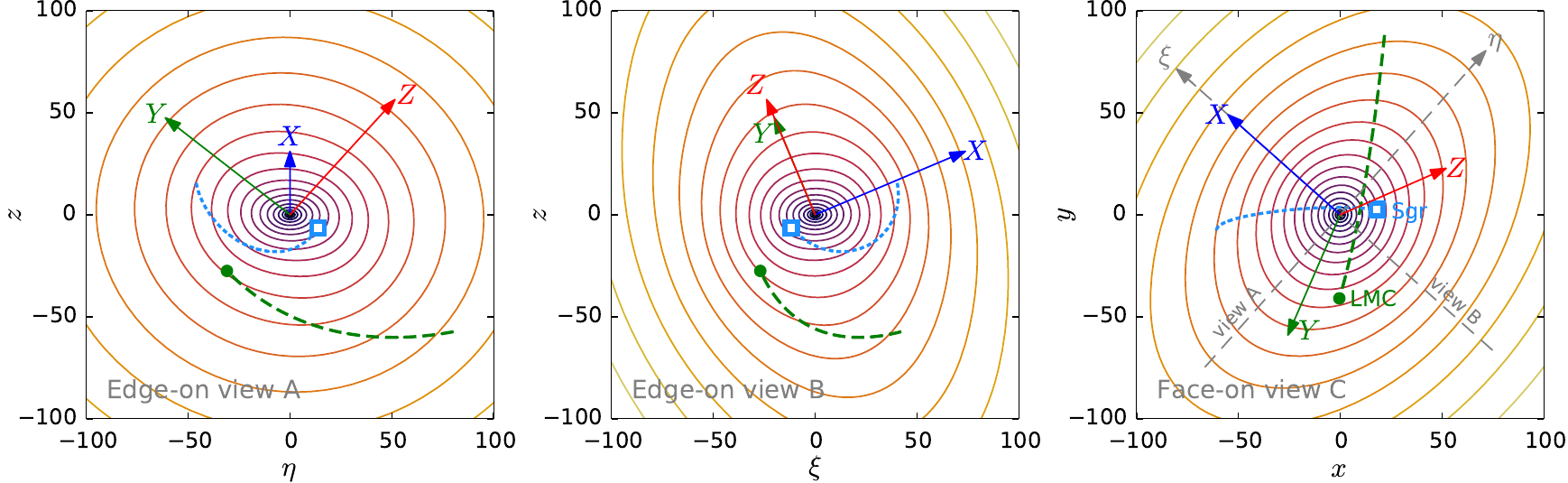}
\caption{Examples of the Milky Way halo shapes from the series of models without the LMC (\textit{top row}) and with the LMC mass $1.5\times10^{11}\,M_\odot$ (\textit{bottom row}). 
Shown are cross-sections of the density in the three perpendicular planes: the Milky Way disc plane (\textit{right} panels, aligned with the standard Galactocentric coordinates) and two edge-on views along the auxiliary axes $\eta$ (\textit{left}) and $\xi$ (\textit{centre}); these axes are shown in the right panel by gray dashed lines. Three coloured arrows show the orientation of the principal axes of the outer halo. Isolines of constant 3d density are spaced at 4 levels per decade. The current position and the trajectory of the LMC over the last 0.5~Gyr are shown by a green dashed line ending with a dot; the current position of the Sgr and its past trajectory -- by a cyan dotted line ending with a square.\protect\\
In both cases, the halo is oblate with an axis ratio $R:z \simeq 2:1$ and aligned with the disc in the inner part, while becoming triaxial and misaligned with the disc in the outer part (beyond $\sim 50$~kpc). In the no-LMC case (a very similar model to the one shown on Figure~\ref{fig:stream_model_nolmc}), the outer part is strongly prolate, with axis ratios $X:Y:Z \simeq 1.5:1:3$. The top left panel shows this twist in the plane A containing the shortest $z$ and the longest $Z$ axes of the inner and the outer halo, respectively (its abscissa $\eta$ is the projection of $Z$ onto the disc plane). In the case with the LMC, the outer part of the halo is close to oblate, with axis ratios $X:Y:Z \simeq 1:1.8:1.8$. The twist is most apparent in the plane B, shown in the bottom centre panel, which contains the shortest axes $z$ and $X$ of the inner and the outer halo, respectively (its abscissa $\xi$ is the projection of $X$ onto the disc plane). This model is a typical example taken from the MCMC series, \textit{not} the fiducial model shown in Figure~\ref{fig:stream_model} (the latter has a very similar orientation of $\xi$, but smaller axis ratio and is not twisted, hence its $Z$ axis points vertically and both $X$ and $Y$ axes lie in the disc plane). See \url{https://youtu.be/F_E0ziJkUPk} and
\url{https://youtu.be/zBtrPk18HYQ} for a 3D view of the isodensity surfaces of the inferred Milky Way dark matter halo with and without the LMC respectively.
} \label{fig:halo_twisted}
\end{figure*}

\begin{figure*}
\includegraphics{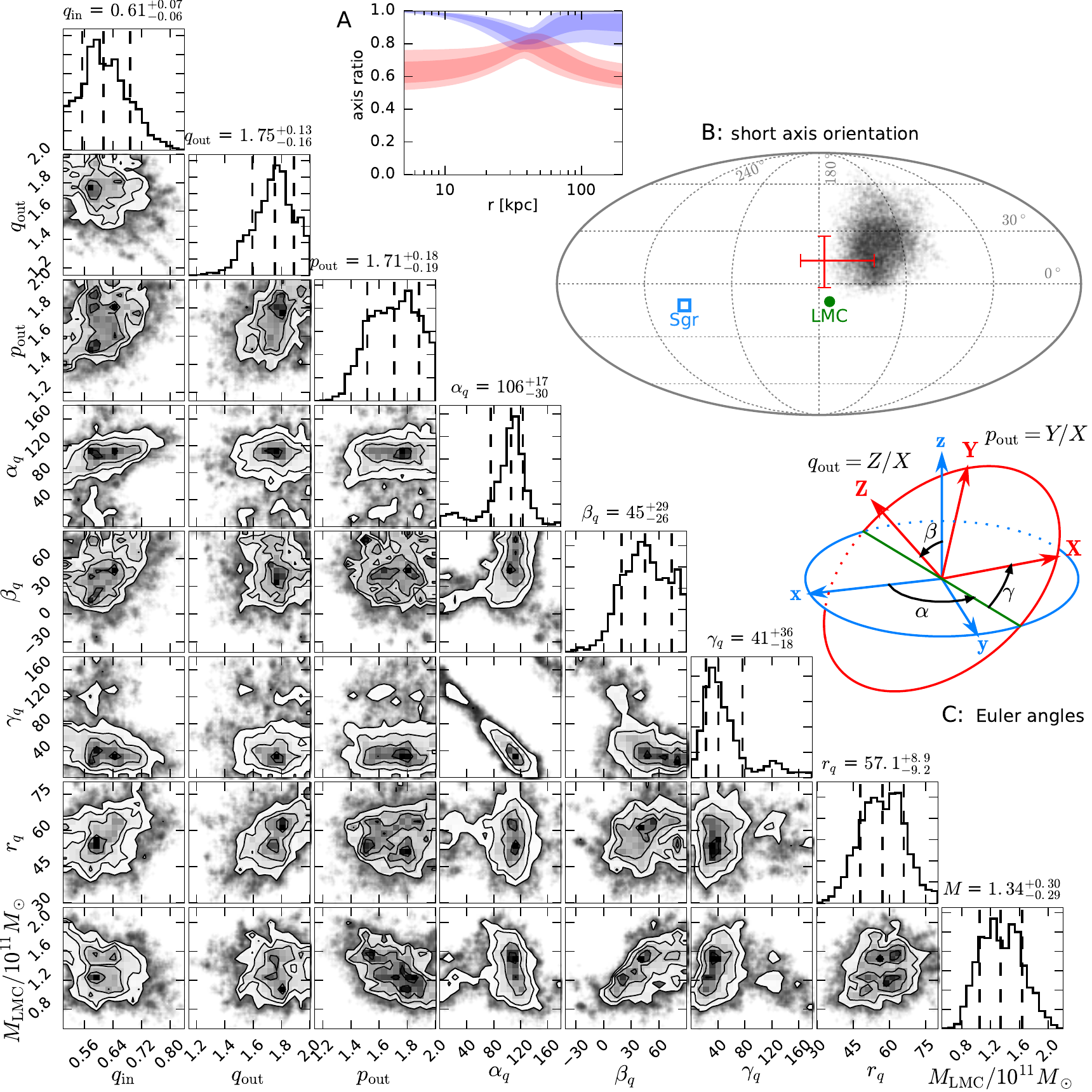}
\caption{Corner plot of Milky Way halo shape parameters from the MCMC run with a variable LMC mass. The inner part of the halo is axisymmetric with the axis ratio $q_\mathrm{in}=z/x$ below unity, i.e.\ oblate. The outer part is triaxial, with the principal axes $X,Y,Z$ rotated w.r.t.\ the Galactocentric coordinate system $x,y,z$ by three Euler angles $\alpha_q, \beta_q, \gamma_q$, as illustrated on the inset panel C. The axis ratios of the outer halo $p_\mathrm{out}=Y/X$ and $q_\mathrm{out}=Z/X$ are both above unity, which means that $X$ is the shortest axis. The transition occurs smoothly around the radius $r_q$ (measured in kpc). Note that there are some geometric degeneracies between parameters: for instance, when the inclination $\beta_q$, which determines the misalignment angle between the $z/Z$ axes of the inner and the outer halo, is small, the two remaining angles $\alpha_q$ and $\gamma_q$ only matter as a sum, not separately. Likewise, when the axis ratio $p_\mathrm{out}$ is close to unity, the angle $\gamma_q$ also doesn't matter. The inset panel A shows the radial profiles of the intermediate to long (blue) and short to long (red) axis ratios; note that the axes labels are swapped around the transition radius. The inset panel B shows the direction of the shortest axis of the outer halo ($X$) in Galactocentric spherical coordinates as a cloud of points around $l\sim 140^\circ$, $b\sim 20^\circ$; the centre of this plot corresponds to the direction from the Galactic centre towards the Sun. For reference, the directions of orbital poles of Sgr and LMC are shown by a blue square and a green dot, respectively. The orientation of the normal to the plane of satellites \citep{Kroupa2005} also approximately coincides with the LMC orbital pole. Red cross shows the inferred orientation of the minor axis in the oblate halo models from \citet{Erkal2019}, who fitted the Orphan stream in the presence of the LMC.
} \label{fig:cornerplot_shape}
\end{figure*}

\begin{figure}
\includegraphics{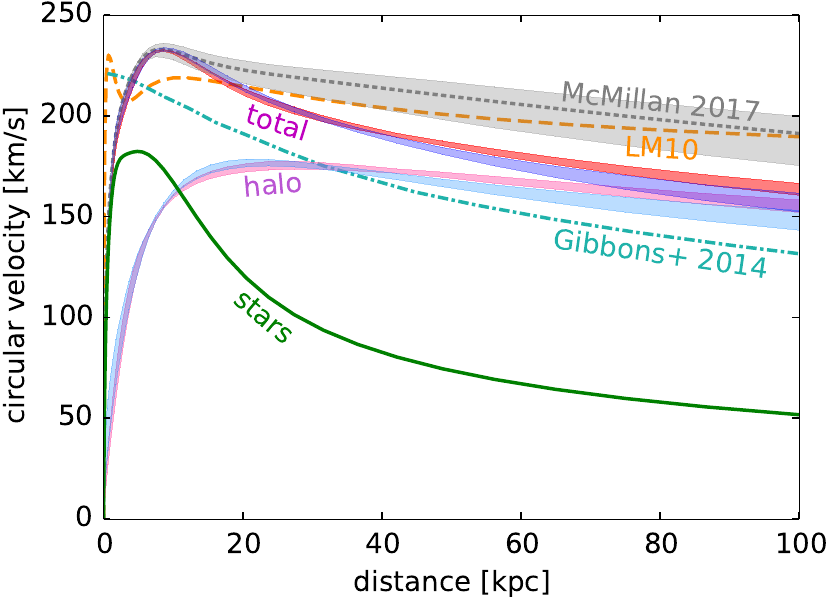}
\caption{
Milky Way circular velocity curve $v_\mathrm{circ} \equiv \sqrt{R\,\partial\Phi/\partial R}$. Solid green line shows the contribution of stars (bulge + disc, same in all models), shaded red and blue regions -- halo and total potential. Models without the LMC are shaded red and have slightly higher velocity, while models with a variable LMC mass in the range $(1-1.7)\times10^{11}\,M_\odot$ are shaded blue and are slightly lower. For comparison, we also plot the circular velocity curves from \citetalias{Law2010a} (dashed orange), \citet{Gibbons2014} (dot-dashed cyan), and \citet{McMillan2017} (dotted gray line is his ``best-fit'' potential, and the $1\sigma$ uncertainty range is shown by a gray shaded band). We used the latter potential as a reference to anchor the circular velocity curve in the inner $\sim10$~kpc and to fix the stellar contribution, but our haloes are much less massive than in that model.
} \label{fig:vcirc}
\end{figure}

Since the Sgr stream is traced over more than $360^\circ$ on the sky and over the range of Galactocentric distances from 20 to 100 kpc, it has long been recognized as a powerful probe of the Milky Way potential. Of course, the large mass of the progenitor and the strong deviation of the stream from the orbit of the Sgr galaxy make this analysis more difficult, and the influence of the LMC adds another degree of complexity (along with extra free parameters pertaining to the LMC itself).

As discussed in Section~\ref{sec:no_LMC}, in absence of the LMC, our best-fit models prefer twisted halo shapes, which change from oblate in the inner part to prolate and misaligned with the disc in the outer part (an example is shown in the top row of Figure~\ref{fig:halo_twisted}). The oblateness of the inner halo is not unreasonable to expect, given the strongly flattened gravitational potential of the stellar and gas discs, which may induce some degree of flattening in the halo due to adiabatic contraction. Furthermore, since we keep the parameters of the stellar potential fixed to some fiducial values, the inferred flattened halo may indicate that the disc mass needs to be revised upward to provide a larger contribution to the overall flattened potential. For comparison, the stellar disc in the \citetalias{Law2010a} model has a mass of $10^{11}\,M_\odot$, and this value is roughly twice higher than the current estimates (see, e.g., \citealt{BlandHawthorn2016}).

The outer halo in the series of LMC-less models is elongated approximately in the direction where the LMC has been a few hundred Myr ago, which hints to the possible explanation of this inferred shape as an artifact mimicking the actual influence of the LMC (see Figure~\ref{fig:halo_twisted}). Likewise, the halo of the \citetalias{Law2010a} model was almost oblate and approximately aligned with the LMC orbital plane (since they forced one of the principal axes to point along the Galactic $z$ axis, their halo could not be twisted in the same way as here). Nevertheless, as stressed above, the moving LMC and the associated reflex motion of the Milky Way cannot be fully emulated by any static potential. Interestingly, we could not obtain good fits when using non-twisted halo models (i.e., fixing $\beta_\mathrm{out}=0$ in Equation~\ref{eq:halo_orientation}). The parameters of the fiducial model shown in Figure~\ref{fig:stream_model_nolmc} are listed in the first column of Table~\ref{table:model_params}.

\begin{table}
\caption{Parameters of the Milky Way halo described in Section~\ref{sec:MW_model} for the fiducial model without the LMC (shown in Figure~\ref{fig:stream_model_nolmc}), with $M_\mathrm{LMC} = 1.5\times10^{11}\,M_\odot$ and an axisymmetric Milky Way halo, and a very similar model with a triaxial outer halo (shown in Figure~\ref{fig:stream_model}). Angles are in degrees, distances -- in kpc; angle $\gamma_q$ is unused (set to 0).
}  \label{table:model_params}
\begin{tabular}{lrrr}
parameter                               & no LMC & axisym. & triax. \\
\hline
scale radius $r_h$                      & 4.20   & 7.00    & 7.36   \\
inner slope $\gamma$                    & 0.45   & 1.00    & 1.20   \\
outer slope $\beta$                     & 2.27   & 2.50    & 2.40   \\
transition steepness $\alpha$           & 2.43   & 2.00    & 2.40   \\
inner $z:R$ axis ratio $q_\mathrm{in}$  & 0.56   & 0.60    & 0.64   \\
outer $Z:X$ axis ratio $q_\mathrm{out}$ & 1.93   & 1.30    & 1.45   \\
outer $Y:X$ axis ratio $p_\mathrm{out}$ & 0.64   & 1.00    & 1.37   \\
angle between $x$ and $X$ $\alpha_q$    & $-44$  &    0    & $-25$  \\
outer inclination angle $\beta_q$       &   52   &    0    &    0   \\
shape transition radius $r_q$           &   69   &   45    &   54
\end{tabular}
\end{table}

In the more realistic models which do take the LMC into account, the inner halo is still oblate approximately in the same way, but the outer part is only moderately non-spherical. We considered several different classes of models. In the first one, the halo is kept axisymmetric, but its outer axis ratio $z:R$ changes from oblate ($q_\mathrm{in}<1$) to mildly prolate ($q_\mathrm{out}\sim 1.2-1.4$). Typical parameters of this series of models are given in the second column of Table~\ref{table:model_params}. These models fit the observations much better than those without the LMC, but the stream track in the leading arm is less strongly bent ``downward'' (towards positive $\Beta$) than the data suggest. Thus the second class of potentials dispenses with the restriction to axisymmetry, and allows the outer halo to have an arbitrary axis ratio, while still keeping its principal axes $X,Y$ within the disc plane (otherwise we would be unable to create a live equilibrium Milky Way model and simulate its encounter with a live LMC at stage B). With the help of two extra free parameters $p_\mathrm{out}$, $\alpha_q$, it is possible to improve the fit to the stream track. The preferred axis ratios are $q_\mathrm{out}\simeq p_\mathrm{out}\sim 1.5$, i.e., the outer halo is nearly oblate, but its short axis lies in the disc plane and points approximately in the direction of the orbital pole of the LMC -- a very similar configuration to \citetalias{Law2010a}, except that the inner halo is flattened and aligned with the disc, rectifying the stability problems pointed out by \citet{Debattista2013}. The fiducial model shown in Figures~\ref{fig:stream_model} and \ref{fig:remnant_model} belongs to this class, and its parameters are listed in the last column of Table~\ref{table:model_params}.

Finally, we explored an even more general class of twisted haloes (same as in the no-LMC case) with restricted $N$-body simulations only. Figure~\ref{fig:cornerplot_shape} shows the range of model parameters related to the halo shape and orientation from the MCMC run, and the bottom row of Figure~\ref{fig:halo_twisted} shows a typical example. Although it may be rather difficult to grasp the details, in general the preferred outer halo shape is approximately oblate, with two axes ($Y$ and $Z$) roughly equal, and the third ($X$) $\sim1.5-2\times$ shorter. The ratio of long to short axes is somewhat larger in this series of models than in the previous (triaxial, non-twisted) case, but the improvement in log-likelihood from allowing a more general orientation is moderate ($\lesssim 5$). The direction of the minor axis is now within $30-40^\circ$ from the orbital pole of the LMC (we remind that the LMC and Sgr orbital planes are close to the $y-z$ and $x-z$ planes, respectively, while the Milky Way disc lies in the $x-y$ plane). It is unclear whether this alignment is still an indication of some deficiency in treating the dynamical influence of the LMC, or could indicate a real physical effect. \citet{Erkal2019} explored constraints on the Milky Way potential from the Orphan stream, also taking the LMC into account and allowing for a general shape and orientation of the halo (although with a radially constant flattening in the potential rather than density, and without a twist); these fits also preferred an asphericity in the halo and found oblate configurations approximately aligned with the LMC orbital plane (shown by a red cross in panel B of Figure~\ref{fig:cornerplot_shape}), consistent with the orientation found in this work. 

The Sgr fits presented in \citet{Cunningham2020} also find a broadly similar halo. In that work, they use a generalized triaxial dark matter halo from \citet{Bowden2013} while also accounting for the LMC. The best-fit halo has a short axis in the direction of $(l_{\rm GC}, b_{\rm GC}) = (0.7^\circ,13.6^\circ)$, broadly consistent with our halo orientation. The inner axes have flattenings of $q_0=0.68, p_0 = 0.87$ and the outer halo flattenings of $q_\infty = 0.81, p_\infty = 0.94$. The best-fit LMC mass is $2.0\times10^{11} M_\odot$. We note that the potential used in that work has less flexibility since the orientation of the halo does not change with radius. In addition, the density of the dark matter halo was taken to be NFW while the potential used in this work has much more flexibility in its profile (equation \ref{eq:halo_density}). Although not shown in \citet{Cunningham2020}, we note that the Sgr model presented in this work provides a much better match to the distances of Sgr stars, likely due to the flexibility of the potential. 

Radially varying shapes and orientations are common in cosmological $N$-body simulations: e.g., \citet{Emami2020} examined Milky Way analogues in the Illustris-TNG50 simulation and found $\sim1/3$ of haloes to have a gradual tilt in the direction of the short axis by up to $90^\circ$, while in another $\sim1/3$ the roles of short and long axes are swapped at some radius -- just as in our models (panel A in Figure~\ref{fig:cornerplot_shape}). Intriguingly, \citet{Shao2020} recently found similar trends in the halo geometry (oblate outer halo with a minor axis lying close to the LMC orbital pole) for a sample of Milky Way analogues in the EAGLE simulations, selected by the condition that their satellites preferentially lie in a single plane, as happens to be for the Milky Way (see e.g.\ \citealt{Kroupa2005,Pawlowski2020}). Nevertheless, we caution against over-interpreting the results of our fits, since the inference about the halo shape is rather sensitive to the parameters of the LMC and the Sgr progenitor, and it is plausible that further variation of these parameters may change the preferred orientation. Since we used a considerable freedom in choosing the halo shape to fit the stream, it would be interesting to see if the models based on modified gravity could reproduce its properties while varying only the parameters of baryons \citep[e.g.,][]{Thomas2017}.

Figure~\ref{fig:vcirc} shows the range of possible Milky Way circular velocity curves for the models without the LMC and with a variable LMC mass in the range $(1.3\pm0.3)\times10^{11}\,M_\odot$. These are taken from the MCMC runs performed with the restricted $N$-body simulation method (Section~\ref{sec:fast_stream_generation}), unlike Figures~\ref{fig:stream_model_nolmc}--\ref{fig:remnant_model}, which show the more realistic full $N$-body models. Nevertheless, the basic trends in the potential are quite robust and can be assessed even from these simpler models. In models without the LMC, the circular velocity is slightly higher in the outer parts -- an effect also noticed by \citet{Erkal2020a} in the context of fitting equilibrium models to the Milky Way stellar halo. Our circular velocity curves at large radii are roughly halfway between those inferred by \citetalias{Law2010a} and \citet{Gibbons2014} from the same Sgr stream. Although the halo profiles are non-spherical and have a rather general functional form (Equation~\ref{eq:halo_density}), the halo circular velocity curve is reasonably well approximated by a simple NFW model with a scale radius 12~kpc and a normalization of 150~\kms at 100~kpc. The scale radius is considerably lower than expected from dark matter-only cosmological simulations, however, the subsequent baryonic contraction would have increased the halo concentration, resulting in a smaller scale radius if we stick to the familiar NFW parametrization \citep[e.g.,][section 5.3.2]{Cautun2020}.

The total enclosed mass within 50~kpc is $(3.85\pm0.1)\times10^{11}\,M_\odot$, and within 100~kpc -- $(5.7\pm0.3)\times10^{11}\,M_\odot$ in our preferred models including the LMC. These values are comfortably in the range inferred by various other studies (see, e.g., figure~2 in \citealt{Wang2020} for a recent compilation). For comparison, \citet{Gibbons2014} measured $(4.0\pm0.7)\times10^{11}\,M_\odot$ within 100~kpc, and \citet{Fardal2019} -- $(7.1\pm0.7)\times10^{11}\,M_\odot$. Although the mass profile is tightly constrained only within the spatial extent of the Sgr stream, we also obtain rather precise yet conservative estimates of the total Milky Way mass from our very flexible parametrization of the halo density profile at large radii (a power law with an exponential cutoff of adjustable sharpness). We use the standard definition of $r_\Delta$, $M_\Delta$ as the radius and corresponding enclosed mass of the spherical region with a mean density equal to $\Delta_\mathrm{c}\,\rho_\mathrm{c}$, where $\rho_\mathrm{c}\equiv 3H^2/8\pi G \simeq 135\,M_\odot\,\mbox{kpc}^{-3}$ is the critical density of the Universe (adopting value of the Hubble parameter $H=70\,\mbox{km\,s}^{-1}\,\mbox{Mpc}^{-1}$). For the values of overdensity $\Delta_\mathrm{c}=102$ \citep[corresponding to the standard definition of the virial mass/radius,][]{Bryan1998} or 200, we obtain $M_\mathrm{vir} = (9.0 \pm 1.3) \times 10^{11}\,M_\odot$ and $r_\mathrm{vir} = 250\pm 12$~kpc, or $M_{200} = (8.0 \pm 1.0) \times 10^{11}\,M_\odot$ and $r_{200} = 192\pm 8$~kpc.

\section{Conclusions}  \label{sec:conclusions}

The Sgr stream provides a rich source of information about the Milky Way system, but is notoriously difficult to model. In this paper, we explored yet another aspect of this complexity arising from the interaction between the Milky Way and the LMC, which creates a significant time-dependent perturbation in the kinematics of all tracers in the halo -- including the Sgr stream. We have uncovered a smoking-gun observational evidence of the time-dependent potential manifested as the misalignment between the stream track and the proper motions of its stars, which is not possible to explain in a static potential. While the encounter with the LMC and the associated reflex motion of the Milky Way is the most natural mechanism generating this time-dependent perturbation, other possibilites are worth examining (e.g., a tumbling Milky Way halo, \citealt{Valluri2020}).

We developed two novel techniques for modelling the Sgr stream in the presence of the LMC. The first one (restricted $N$-body simulations -- stage A) is conceptually similar to various fast stream generation approaches used in previous papers \citep{Gibbons2014, Fardal2019}, but slightly more sophisticated, since the particle orbits are followed in a realistic moving potential of the progenitor and escape through Lagrange points in a natural way, instead of being placed there artificially. This method is fast enough to allow an MCMC exploration of the parameter space of the Milky Way potential and identify the plausible range of parameters. However, the resulting stream properties are slightly different from those obtained by fully self-consistent $N$-body simulations, which are the other cornerstone method of this study. We use a two-stage approach to simulate the interaction of the three galaxies: first run a live Milky Way + LMC simulation and represent it in a condensed way with basis-set expansions of the evolving potentials of both galaxies (stage B), then explore the disruption of the Sgr galaxy in this pre-recorded combined potential as a live $N$-body system (stage C). These simulations are far more expensive, but allow us to obtain a very realistic representation of the stream for selected potential parameters taken from the pool of MCMC models with high likelihood.

We demonstrated that the models with a static but flexible Milky Way halo parametrization but no LMC can fit most of the observed stream properties quite well, although they require a rather peculiar shape and orientation of the halo, which is markedly prolate in the outer part. However, models from this series fail to reproduce two aspects of the stream: the apocentre distance of the leading arm is overestimated, and there is no misalignment between the stream track and its proper motion. 

On the other hand, models with the LMC of mass $M_\mathrm{LMC}=1.5\times10^{11}\,M_\odot$ alleviate these deficiencies at least qualitatively, while adequately matching various other stream features.  A similar LMC mass is derived from fitting the Orphan stream \citep{Erkal2019}, which also shows the same misalignment in one of its branches. Other streams in the southern hemisphere display various levels of misalignment, depending on the relative position of the stream and the LMC \citep[][]{Shipp2019,Li2020}. A joint analysis of multiple streams may shed light on the mass distribution of the LMC (Shipp et al., in prep.). At the same time, we found that the modelled Sgr stream is affected by the distortions in the LMC and the Milky Way halo in a live simulation of their encounter -- another level of complexity that may be used to constrain the properties of both systems (Lilleengen et al., in prep.).

Even in the presence of the LMC, our model fits prefer a non-spherical Milky Way halo that changes shape with radius from oblate and aligned with the disc in the inner part to still oblate but almost perpendicular to the disc in the outer part. The axis ratio of the outer halo density lies in the range $1.5-2$, and the shortest axis direction lies within $30^{\circ}-40^\circ$ from the orbital pole of the LMC (although both these parameters have large uncertainties and can depend on other model assumptions). On the other hand, the constraints on the radial profile of the halo, manifested in the circular-velocity curve, are rather tight and insensitive to the details of the models; the enclosed mass within 100~kpc is estimated as $(5.6\pm0.4)\times10^{11}\,M_\odot$ and agrees comfortably with a number of independent recent measurements.

The motion of the Milky Way in response to the in-fall of the LMC plays a crucial role in our models. We show that the action of a massive Cloud in a static host can lead to a strong perturbation of the Sgr stream, but its direction appears to be opposite to that needed to reproduce the stream track misalignment discovered here. Conversely, the additional accelerations in the rebounding non-inertial Milky Way help to reshape the stream in 3d and to give its leading tail the necessary amount of sideways motion. We therefore conclude that the observed structure and kinematics of the Sgr stream provides one of the strongest proofs of the ongoing recoil of our Galaxy. 

Our models inevitably have a number of shortcomings and limitations:
\begin{itemize}
\item We have primarily focused on spherical, non-rotating Sgr progenitors. The observed properties of the remnant are sensitive to the amount of residual rotation, and indeed a small amount of it may be needed to better match the line-of-sight velocity field and the angle between the stream track and the major axis of the remnant, which is around $4^\circ$ in the data but only $2^\circ$ in the model. Strong rotation, however, would be in tension with observed properties of the remnant, as shown by \citetalias{Vasiliev2020}.
\item The observed bifurcation in the stream is not reproduced by models. It might possibly be explained by a superposition of rotating and non-rotating components in the progenitor (e.g., \citealt{Penarrubia2010}, although that particular model was not supported by later observations).
\item The stream track in the leading arm is offset by $\sim2^\circ$ in our models from the observed track. The dependence of $\Beta$ on $\Lambda$ is sensitive to the shape of the Milky Way halo, but also to the rotation of the progenitor \citep{Gibbons2016}, making it difficult to disentangle the role of these two factors without running a representative number of models. Because of these uncertainties, we regard our inference on the halo shape as preliminary.
\end{itemize}

Despite these caveats, our models provide an important stepping stone to understanding the entangled dance of the three stellar systems, but of course are far from the ultimate truth.  Indeed, the Sgr stream is as inexhaustible as the atom, and explaining its various features will keep theorists busy for years to come.
 
\section*{Acknowledgements}

We thank Larissa Palethorpe for carrying out the analysis of globular cluster association with the Sgr stream, and Sergei Koposov and other colleagues for valuable discussions. EV acknowledges support from STFC via the Consolidated grant to the Institute of Astronomy.  This work uses the data from the European Space Agency mission \Gaia (\url{https://www.cosmos.esa.int/gaia}), processed by the \Gaia Data Processing and Analysis Consortium (\url{https://www.cosmos.esa.int/web/gaia/dpac/consortium}). 

VB recollects the talk by S. Majewski at the ``Mass and Mystery in the Local Group'' conference (2005) in Cambridge to commemorate the 70th birthday of Professor Donald Lynden-Bell. In his presentation, Steve conjectured that a triple interaction between the Milky Way, the LMC and the Sagittarius dwarf may be needed to explain the stream behaviour. At the time, this idea appeared very brave, even wild. We needed to wait 15 years for the marvelous \Gaia data to be able to test this hypothesis.

\section*{Data availability} 

We provide the catalogue of $\sim55\,000$ high-probability stream members described in Section~\ref{sec:data}, which combines the positions and proper motions from \Gaia, distances estimated from \Gaia RR Lyrae stars, and for a small subset of stars ($\sim 4\,500$) -- line-of-sight velocities from various spectroscopic surveys. This catalogue is hosted at \url{https://zenodo.org/record/4038137} and will also be published on CDS and as the supplementary online material.

We publish the fiducial model of the Sgr stream in the presence of a $1.5\times10^{11}\,M_\odot$ LMC as a snapshot of the $N$-body simulation at the present time (stellar and dark matter components). We also provide the simulation of the Milky Way + LMC encounter for this model, represented by time-dependent multipole expansions of their potentials, trajectories of Sgr and LMC in the Milky Way-centered reference frame, and the acceleration of this non-inertial frame. These data can be used to explore the effect of the LMC on various other kinematic tracers in the Milky Way halo -- streams, stars, clusters, etc. Finally, we include Python scripts for constructing initial conditions for these simulations and for integrating orbits in a time-dependent potential, using the \textsc{Agama} framework\footnote{Available at \url{http://agama.software}}. The archive with these data is hosted at \url{https://zenodo.org/record/4038141}.


\appendix
\section{Restricted vs full $N$-body models} \label{appendix:compare}

\begin{figure*}
\includegraphics{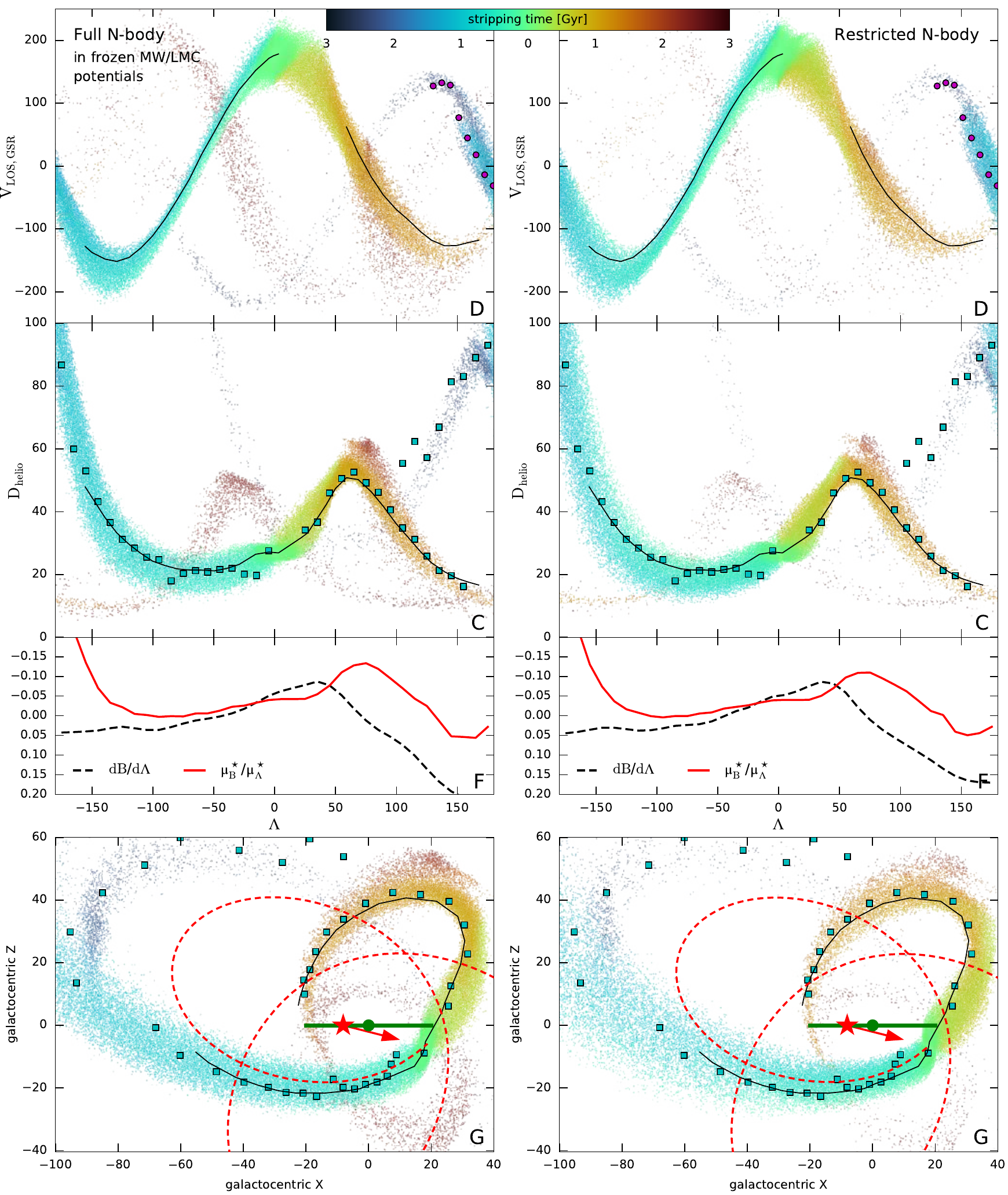}
\caption{Comparison of full (left column) and restricted (right column) $N$-body simulations for the Sgr stream model including the LMC. Both cases use frozen (non-evolving) Milky Way and LMC potentials, unlike the full $N$-body simulation shown in Figure~\ref{fig:stream_model}, which used deforming potentials of the Milky Way halo and the LMC. Shown are a subset of panels from that figure: line-of-sight velocity (first row), heliocentric distance (second row), misalignment between the stream track and the proper motion (third row), and Galactocentric side-on view (bottom row). Although the restricted $N$-body simulation underpredicts the number of particles stripped at early times, the agreement in overall stream properties between the two approaches is otherwise fairly good.
} \label{fig:stream_compare_restricted}
\end{figure*}

Figure~\ref{fig:stream_compare_restricted} shows a comparison of streams generated by the restricted $N$-body simulation approach (stage A, described in section~\ref{sec:fast_stream_generation}) and live $N$-body simulations of a disrupting satellite in ``frozen'' (non-evolving) Milky Way + LMC potentials (stage C). The morphological and kinematic properties of the stream are well reproduced by the faster restricted simulation approach, justifying its usage in the MCMC exploration of the parameter space. However, we remind that the live $N$-body simulations with evolving Milky Way and LMC potentials (represented by multipole expansions extracted from a separate $N$-body simulation in stage B) do show subtle but noticeable differences from simulations in a ``frozen'' potential (the latter jump directly to stage C, omitting stage B).

\end{document}